\pdfoutput=1
\documentclass[letterpaper,11pt]{article}

\usepackage{jheppub}
\usepackage{graphicx}
\usepackage{amsmath}
\usepackage{amssymb}
\usepackage{mathrsfs}
\usepackage{xspace}

\newcommand{\eq}[1]{eq.~\eqref{eq:#1}}
\newcommand{\eqs}[2]{eqs.~\eqref{eq:#1} and \eqref{eq:#2}}
\renewcommand{\sec}[1]{sec.~\ref{sec:#1}}
\newcommand{\secs}[2]{secs.~\ref{sec:#1} and \ref{sec:#2}}

\newcommand{\app}[1]{app.~\ref{app:#1}} 
\newcommand{\fig}[1]{fig.~\ref{fig:#1}}

\newcommand{\abs}[1]{\lvert#1\rvert}
\newcommand{\Abs}[1]{\bigl\lvert#1\bigr\rvert}
\newcommand{\ord}[1]{{\mathcal O}(#1)}
\newcommand{\ORd}[1]{{\mathcal O}\Bigl(#1\Bigr)}

\newcommand{\MAe}[3]{\Bigl\langle#1\Bigr\rvert#2\Bigr\rvert#3\Bigr\rangle}

\newcommand{\nn}{\nonumber}

\newcommand{\df}{\mathrm{d}}
\newcommand{\img}{\mathrm{i}}
\newcommand{\sdt}{\!\cdot\!}

\newcommand{\lra}{\leftrightarrow}

\newcommand{\al}{\alpha}
\newcommand{\bt}{\beta}
\newcommand{\ga}{\gamma}
\newcommand{\Ga}{\Gamma}
\newcommand{\de}{\delta}

\newcommand{\eps}{\epsilon}

\newcommand{\la}{\lambda}
\newcommand{\si}{\sigma}

\newcommand{\cI}{{\mathcal I}}

\newcommand{\cL}{{\mathcal L}}

\newcommand{\cS}{{\mathscr S}}

\newcommand{\bnslash}{\bar{n}\!\!\!\slash}

\newcommand{\vks}{\vec k_\perp^{\,2}}
\newcommand{\vkps}{\vec k_\perp^{'2}}

\newcommand{\bn}{{\bar{n}}}

\newcommand{\lqcd}{\Lambda_\mathrm{QCD}}

\newcommand{\cusp}{\mathrm{cusp}}

\newcommand{\Tau}{{\mathcal T}}

\newcommand{\zero}{{(0)}}
\newcommand{\one}{{(1)}}

\newcommand{\SCETa}{\ensuremath{{\rm SCET}_{\rm I}}\xspace}
\newcommand{\SCETb}{\ensuremath{{\rm SCET}_{\rm II}}\xspace}
\newcommand{\SCETab}{\ensuremath{{\rm SCET}_+}\xspace}

\newcommand{\CS}{\cS}

\newcommand{\Ecm}{E_\mathrm{cm}}

\allowdisplaybreaks[2]

\title{Resummation of Double-Differential Cross Sections and Fully-Unintegrated Parton Distribution Functions}

\author[a]{Massimiliano Procura,}
\author[b,c]{Wouter J.~Waalewijn,}
\author[b]{Lisa Zeune}

\affiliation[a]{Albert Einstein Center for Fundamental Physics, Institute for Theoretical Physics, \\ University of Bern, CH-3012 Bern, Switzerland}
\affiliation[b]{ITFA, University of Amsterdam, Science Park 904, 1018 XE, Amsterdam, The Netherlands}
\affiliation[c]{Nikhef, Theory Group, Science Park 105, 1098 XG, Amsterdam, The Netherlands}

\emailAdd{mprocura@itp.unibe.ch}
\emailAdd{wouterw@nikhef.nl}
\emailAdd{l.k.zeune@uva.nl}

\abstract{LHC measurements involve cuts on several observables, but resummed calculations are mostly restricted to single variables. We show how the resummation of a class of double-differential measurements can be achieved through an extension of Soft-Collinear Effective Theory (SCET). A prototypical application is $pp \to Z + 0$ jets, where the jet veto is imposed through the beam thrust event shape $\Tau$, and the transverse momentum $p_T$ of the $Z$ boson is measured. A standard SCET analysis suffices for $p_T \sim m_Z^{1/2} \Tau^{1/2}$ and $p_T \sim \Tau$, but additional collinear-soft modes are needed in the intermediate regime. We show how to match the factorization theorems that describe these three different regions of phase space, and discuss the corresponding relations between fully-unintegrated parton distribution functions, soft functions and the newly defined collinear-soft functions.
The missing ingredients needed at NNLL/NLO accuracy are calculated, providing a check of our formalism. We also revisit the calculation of the measurement of two angularities on a single jet in JHEP 1409 (2014) 046, finding a correction to their conjecture for the NLL cross section at $\ord{\al_s^2}$.}

\begin{document}
{\flushright NIKHEF 2014-043 \\[-6ex]}

\maketitle

\section{Introduction}
\label{sec:intro}

Experimental LHC analyses typically involve several kinematic cuts. Many of them are fairly harmless from a theoretical point of view. However, when these restrictions on initial- and/or final-state radiation lead to widely separated energy scales, large logarithms can be induced in the corresponding cross section, requiring resummation. One example is given by the jet veto used to suppress backgrounds in Higgs analyses, where the resummation of jet-veto logarithms~\cite{Berger:2010xi,Banfi:2012jm,Liu:2013hba,Becher:2013xia,Stewart:2013faa,Boughezal:2013oha} greatly reduces the dominant source of theoretical uncertainty. A closely related process is Drell-Yan (or vector boson) production in the case the lepton pair has a small $p_T$ compared to their invariant mass $Q$~\cite{Dokshitzer:1978yd,Parisi:1979se,Collins:1984kg,Davies:1984sp,Altarelli:1984pt,Catani:2000vq,deFlorian:2001zd,Becher:2010tm,Idilbi:2005er,Mantry:2010mk,GarciaEchevarria:2011rb}. Another example is the jet mass $m_J$ spectrum of a jet with transverse momentum $p_T^J$, which requires resummation around the peak of the distribution where $m_J \ll p_T^J$~\cite{Li:2012bw,Dasgupta:2012hg,Chien:2012ur,Jouttenus:2013hs}.

In this paper we focus on double differential measurements, where both observables lead to large logarithms. Using effective field theory methods, we derive new resummed expressions for a class of double differential cross sections. 
Our results smoothly connect to the phase space boundaries, which require different effective field theories.
This formalism has applications to jet cross sections and jet substructure studies, and we will consider an example of both in this paper.

As the field of jet substructure has matured~\cite{Abdesselam:2010pt,Altheimer:2012mn,Altheimer:2013yza}, multivariate analyses have become common. Furthermore, some of the measurements with the best discrimination power are ratios of infrared and collinear safe observables, such as ratios of $N$-subjettiness~\cite{Stewart:2010tn,Thaler:2010tr,Thaler:2011gf}, energy correlation functions~\cite{Banfi:2004yd,Larkoski:2013eya,Larkoski:2014gra} or planar flow~\cite{Thaler:2008ju,Almeida:2008yp}. These quantities are themselves not infrared and collinear safe, and their calculation involves marginalizing over the \emph{resummed} two-dimensional distribution~\cite{Larkoski:2013paa}. The pioneering study in ref.~\cite{Larkoski:2014tva},  investigating the measurement of two angularities on one jet, inspired the present paper. 

Our formalism can also be applied to $p p \to H+1$ jet production, where in addition to the jet veto the transverse momentum of the jet becomes small. This important contribution to the cross section is not yet fully understood~\cite{Boughezal:2013oha}. In this paper, to better illustrate the features of our framework, we will mainly focus on a simpler (but related) problem in $Z+0$ jet production, carrying out the simultaneous resummation of the jet veto and the transverse momentum of the $Z$ boson.

Resummation is often achieved using the parton shower formalism. The great advantage of parton shower Monte Carlo event generators, such as \textsc{Pythia}~\cite{Sjostrand:2006za} and \textsc{Herwig}~\cite{Bahr:2008pv}, is that they produce a fully exclusive final state, giving the user full flexibility. On the other hand, this approach is limited to leading logarithmic (LL) accuracy, and it is difficult to estimate the corresponding theory uncertainty. It is also not clear to what extent correlations between resummed observables are correctly predicted by Monte Carlo models, see e.g.~ref.~\cite{Larkoski:2014pca}. By contrast, we predict these correlations and our resummed predictions have a theory uncertainty attached to it, whose reliability can be verified by comparing different orders in resummed perturbation theory. Note that there has been significant progress by matching higher-order matrix elements with parton showers (see e.g.~refs.~\cite{Frixione:2002ik,Nagy:2005aa,Frixione:2007vw,Alioli:2010xd,Alioli:2012fc,Hamilton:2013fea,Alioli:2013hqa,Hoeche:2014aia}) and (partially) including higher-order resummation~\cite{Alioli:2012fc}.

We will illustrate the features of our framework in the specific case of $pp \to Z+0$ jets, where the transverse momentum $p_T$ of the $Z$ boson is measured and a global jet veto is imposed using the beam thrust event shape~\cite{Stewart:2009yx,Berger:2010xi}
\begin{align} \label{eq:Tau}
\Tau
= \sum_i\, p_{iT}\, e^{-\abs{\eta_i}}
= \sum_i\, \min \{p_i^+,p_i^-\}
\,.\end{align}
The sum on $i$ runs over all particles in the final state, except for the leptonic decay products of the $Z$.
Here, $p_{iT}$ is the magnitude of the transverse momentum and $\eta_i$ the pseudorapidity of particle $i$ in the center-of-mass frame of the hadronic collision. Light-cone coordinates are defined as
\begin{align} \label{eq:lc}
  p_i^\mu = p_i^+ \frac{n^\mu}{2} + p_i^- \frac{n^\mu}{2} + p_{i\perp}^\mu
  \,, \qquad
   p_i^- = \bn \sdt p_i
  \,, \qquad
   p_i^+ = n \sdt p_i
\,,\end{align}
where $n^\mu = (1,0,0,1)$ and $\bn^\mu = (1,0,0,-1)$ are along the beam axis.
Beam thrust imposes a global veto on all radiation in an event, which is impractical in the 
LHC environment. This can be remedied by only including the contributions from jet regions in \eq{Tau}~\cite{Tackmann:2012bt}. We will nevertheless consider the global veto to keep our discussion as simple as possible. At the end of \sec{fact_formula} we will comment on a special class of non-global measurements whose logarithms can easily be resummed within our approach.

 \begin{figure}[t]
  \centering
   \includegraphics[width=0.5\textwidth]{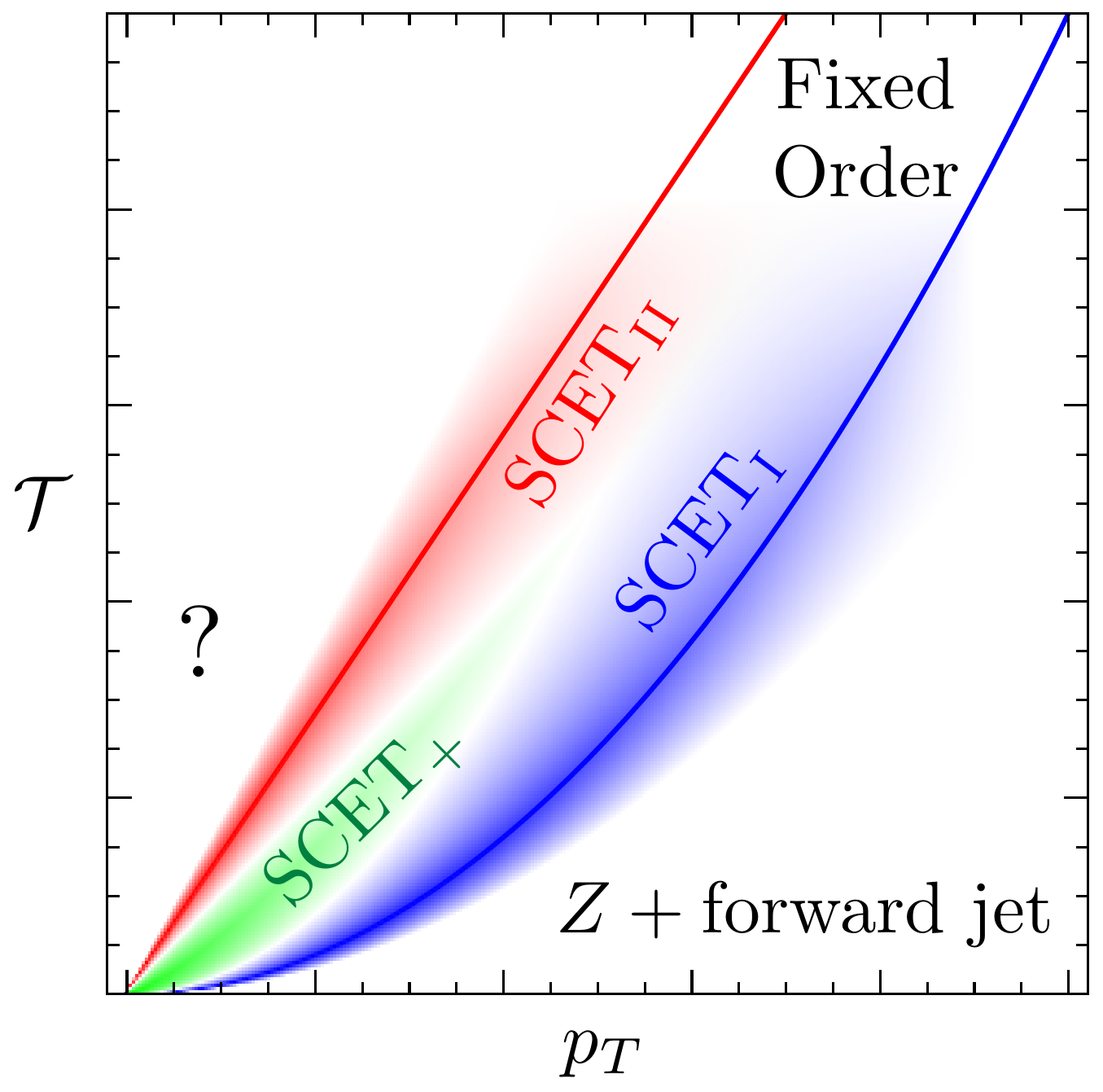}
   \caption{The different regions for the double measurement of $p_T$ and beam thrust $\Tau$ in $Z$-boson production from $pp$ collisions.}
   \label{fig:regions}
  \end{figure}

\begin{figure}[t]
   \includegraphics[width=0.99\textwidth]{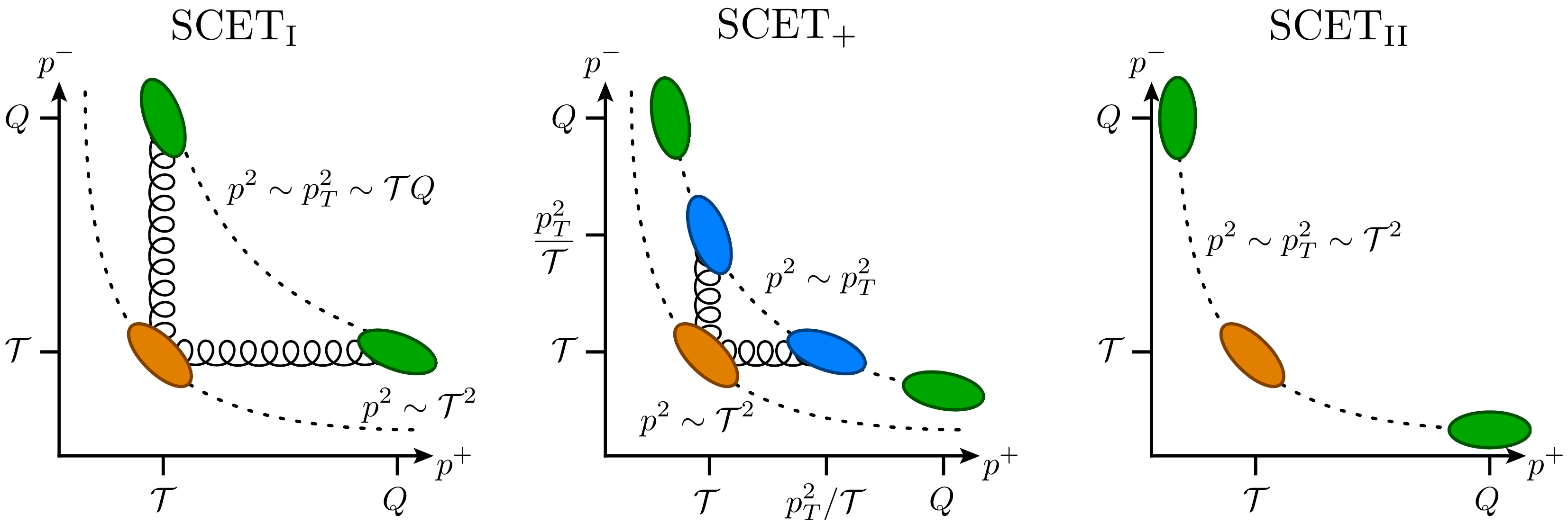}
   \caption{The modes in \SCETa, \SCETab and \SCETb: collinear (green), collinear-soft (blue) and soft (orange). Interactions between modes in the effective theory are shown with wiggly lines. These are removed by the decoupling transformations in \eq{BPS}.}
  \label{fig:modes}
\end{figure}

We will perform resummations using Soft-Collinear Effective Theory (SCET)~\cite{Bauer:2000ew, Bauer:2000yr, Bauer:2001ct, Bauer:2001yt}. Which version of SCET is the appropriate one, namely what the relevant degrees of freedom are, depends on the region of phase space probed by the measurement, as shown in \fig{regions} and discussed below. We find that in the intermediate region, between the \SCETa and \SCETb boundaries, the effective field theory involves additional collinear-soft modes. This type of mode was introduced in a different context in ref.~\cite{Bauer:2011uc}, and has led us to also refer to our effective theory as \SCETab. Since we are considering different observables than ref.~\cite{Bauer:2011uc}, there are of course important differences, which will be discussed in \sec{scet}.
We now comment on the theoretical description relevant for each region of phase space in the $(p_T, \Tau)$ plane.
\begin{itemize}
     \item
     Fixed Order: $p_T,  \Tau\sim Q$\\
     When $p_T$ and $\Tau$ are parametrically of the same size as the hard scale $Q^2 = p_Z^2 \sim m_Z^2$, resummation is not necessary and a fixed-order calculation suffices.
     \item
      \SCETa: $p_T \sim Q^{1/2} \Tau^{1/2}$ \\
      This case was discussed in ref.~\cite{Jain:2011iu}. The collinear and soft modes, shown in the left panel of \fig{modes}, interact. 
      The \SCETa scale hierarchy implies that the soft radiation contributes only to $\Tau$ (its contribution to $p_T$ is power suppressed), whereas the collinear radiation contributes both to the $p_T$ and the $\Tau$ measurement. This collinear radiation is described by fully-unintegrated parton distribution functions (PDFs)~\cite{Collins:2007ph,Rogers:2008jk,Mantry:2009qz,Jain:2011iu},
which depend on all momentum components of the colliding parton. By contrast, the standard PDFs only depend on the momentum fraction $x$.
     \item 
      \SCETab: $p_T \sim Q^{1-r} \Tau^r$ with $1/2<r<1$ \\
      As $p_T$ is lowered, the collinear modes can no longer interact with the soft mode. 
      They ``split off" collinear-soft modes that do interact with the soft modes, see \fig{modes}.
      (To have a distinct mode contribution requires sufficient distance from the \SCETa and \SCETb boundaries.)
      In this scenario, the collinear radiation only contributes to $p_T$, the soft radiation only to $\Tau$, and the collinear-soft radiation enters in both measurements. The \SCETab power counting will be given below in table~\ref{tab:modes}.
     \item
      \SCETb: $p_T \sim \Tau$ \\
      As $p_T$ is reduced further, the soft mode ``absorbs" the two collinear-soft modes. In the resulting theory there are no interactions between the collinear and the soft modes, as shown in the left panel of \fig{modes}. The collinear radiation, which in the \SCETb case is described by transverse-momentum dependent (TMD) PDFs, only affects $p_T$, whereas now the soft mode contributes to both measurements.
     \item 
       $Z+$forward jet: $p_T \gg Q^{1/2} \Tau^{1/2}$  \\       
       As $p_T$ exceeds this bound, the QCD radiation becomes (much) more energetic than the invariant mass $Q$ of the $Z$ boson. This cannot be described as initial-state radiation, but rather as $Z$ production in association with an energetic forward jet. 
     \item 
       Terra incognita:  $p_T \ll \Tau$ \\
       Unlike the previous regions, the cross section no longer receives a contribution from a single emission. There is a small NNLO contribution  from the region of phase space where the two emissions are (almost) back-to-back in the transverse plane. In double parton scattering (DPS) the production of the $Z$ and the two jets are (largerly) independent of each other, causing the jets to naturally be back-to-back.\footnote{This feature is exploited to extract DPS experimentally, see e.g.~refs.~\cite{Myska:2013jna,Chatrchyan:2013xxa}.}
       The contribution from DPS is therefore also important.       
 As the proper method for combining single and double parton scattering is still under debate~\cite{Cacciari:2009dp,Diehl:2011yj,Ryskin:2011kk,Manohar:2012pe,Gaunt:2012dd,Blok:2013bpa}, we leave this for future work.            
\end{itemize}

In this paper, we also show how to combine the \SCETa, \SCETab and \SCETb regions to achieve NNLL resummation throughout. The corresponding next-to-leading order cross section is calculated, providing a check of our results.

In most earlier studies of multi-dimensional observables in SCET, such as refs.~\cite{Ellis:2010rwa,Jouttenus:2011wh}, the measurements concerned different regions of phase space (hemispheres, jets, etc.). There, resummation is achieved by assuming a \emph{single} parametric relation between the observables, to avoid so-called non-global logarithms~\cite{Dasgupta:2001sh,Dasgupta:2002dc}.
In ref.~\cite{Larkoski:2014tva} the two boundary theories for the measurement of two angularities on a single jet were identified. There an interpolating function across the intermediate region was derived, by requiring it to be continuous and have a continuous derivative at the boundaries. We revisit their NLL results and find a discrepancy at $\ord{\al_s^2}$ in the bulk. It is worth mentioning that in this case both boundaries involve \SCETa-type theories, to which our framework can be applied as well.

The paper is structured as follows. In \sec{scet} we introduce \SCETab, perform the matching of QCD onto \SCETab currents, and comment on the (dis)similarities with the theory introduced in ref.~\cite{Bauer:2011uc}. Sec.~\ref{sec:fact_formula} contains the factorization formulae for the Drell-Yan cross section with a simultaneous measurement of $p_T$ and $\Tau$ in the \SCETa, \SCETab and \SCETb regions of phase space, as well as the field-theoretic definitions of the matrix elements involved. We calculate/collect all the ingredients necessary to achieve NNLL accuracy in \sec{ingredients} and discuss the (all-order) matching of \SCETa, \SCETab and \SCETb in \sec{matching}. The corresponding NLO cross section is calculated in \sec{NLO}, providing a verification of our resummed predictions. In \sec{ang} we calculate the double angularity measurement on a single jet and compare with ref.~\cite{Larkoski:2014tva}. Conclusions and outlook are presented in \sec{conclusions}.

\section{Factorization}
\label{sec:fact}

\subsection{Effective Theory for the Region between \SCETa and \SCETb Boundaries}
\label{sec:scet}

Soft-Collinear Effective Theory (SCET)~\cite{Bauer:2000ew, Bauer:2000yr, Bauer:2001ct, Bauer:2001yt} describes the collinear and soft limits of QCD. For a pedagogical introduction see e.g.~refs.~\cite{Stewart:notes,Becher:2014oda}. SCET captures QCD in the infrared regime up to corrections that are suppressed by powers of the SCET expansion parameter $\la \ll 1$, in exchange for enabling the resummation of large logarithms of $\la$. 
As discussed in \sec{intro}, both the process and measurement determine which modes give the leading contributions to the cross section in a specific kinematic regime. In \fig{modes} we summarize the scalings and interactions between different degrees of freedom leading to the physical picture in \sec{intro}. These modes need to be well-separated, in order for $\la$ to be small.
The decoupling of modes in the SCET Lagrangian (at leading power) allows one to factorize multi-scale cross sections into products (or convolutions) of single-scale functions for each mode. At its natural scale, each of these function contains no large logarithms. By applying the renormalization group (RG) evolution from these natural scales to a common scale $\mu$, we achieve resummation of logarithms of $\la$ in the cross section. 
 \begin{figure}[t]
  \centering
   \includegraphics[width=0.35\textwidth]{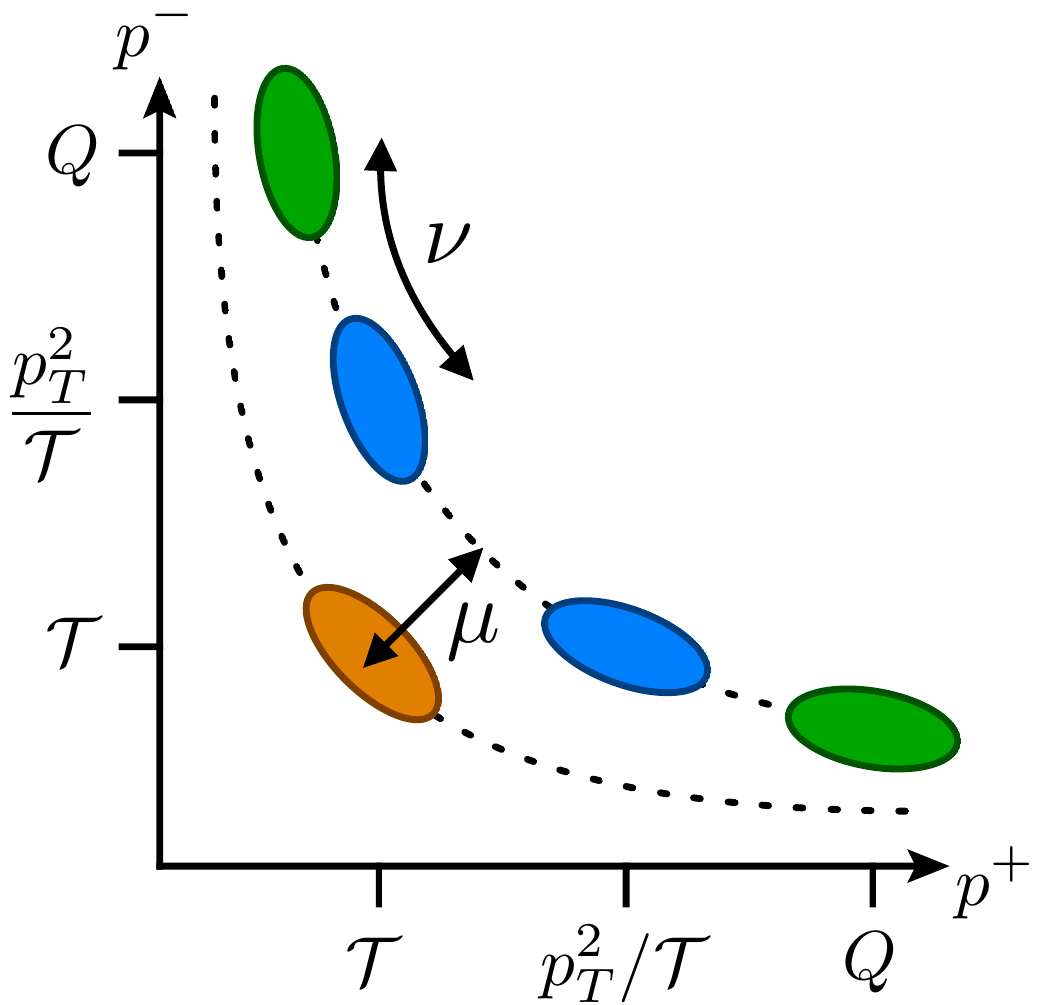}
   \caption{The $\mu$-evolution resums double logarithms from separations in virtuality 
   (between hyperbolae), while the $\nu$-evolution resums single logarithms related to separations in rapidity (along hyperbolae). The collinear, collinear-soft and soft modes are depicted in green, blue and orange, respectively.}
   \label{fig:RGE}
  \end{figure}
For modes that are not separated in virtuality but only in rapidity, we will sum the corresponding single logarithms through the $\nu$-evolution of the rapidity renormalization group~\cite{Chiu:2011qc,Chiu:2012ir}.\footnote{For alternative approaches to rapidity resummation in SCET, see e.g.~refs.~\cite{Chiu:2007dg,Becher:2010tm}.} Pictorially, the $\mu$-evolution sums logarithms related to the separation between the mass hyperbolae of the modes, whereas the $\nu$-evolution sums the logarithms related to the separation along them, see \fig{RGE}.

We will now discuss \SCETab in some detail, focussing on modes, matching of QCD onto \SCETab and factorization.  
We refrain from performing a full formal construction of the effective theory. Factorization means there are no interactions between the various modes, and each mode is described by a (boosted) copy of QCD. In particular, one can use the standard QCD Feynman rules (rather than e.g. the collinear effective Lagrangian of ref.~\cite{Bauer:2000yr}) to carry out the computations for each sector. 

   \begin{table}
   \centering
   \begin{tabular}{l|rcl}
     \hline \hline
     Mode: & Scaling $(- , + ,\perp)$  \\ \hline
     $n$-collinear & $ Q(1,\lambda^{2r},\lambda^r)$&$\sim$&$ (Q,p_T^2/Q,p_T)$  \\
     $\bn$-collinear & $ Q(\lambda^{2r},1,\lambda^r)$&$\sim$&$ (p_T^2/Q,Q,p_T)$ \\
     $n$-collinear-soft & $ Q(\lambda^{2r-1},\lambda,\lambda^r)$ & $\sim$&$ (p_T^2/\Tau,\Tau,p_T)$ \\ 
     $\bn$-collinear-soft & $ Q(\lambda,\lambda^{2r-1},\lambda^r)$ & $\sim$&$ (\Tau,p_T^2/\Tau,p_T)$  \\ 
     soft & $ Q(\lambda,\lambda,\lambda)$ & $\sim$&$ (\Tau,\Tau,\Tau)$  \\
     \hline \hline
   \end{tabular}
   \caption{Modes and power counting in \SCETab with  $\la \sim \Tau/Q \sim (p_T/Q)^{1/r}$.}
   \label{tab:modes}
   \end{table}

   The measurement of beam thrust $\Tau$ and transverse momentum $p_T$, with $p_T \sim Q^{1-r} \Tau^r$ and $1/2<r<1$,\footnote{Note that our analysis is independent of the parameter $r$, as is clear from the second way of writing the modes in table~\ref{tab:modes}. However, we prefer to use a single power counting parameter $\la$.} suggests that the relevant modes are those listed in table~\ref{tab:modes} and shown in the center panel of \fig{modes}, with power counting parameter
   \begin{equation}
    \la \sim \frac{\Tau}{Q} \sim \Big(\frac{p_T}{Q}\Big)^{1/r} 
  \,.\end{equation}
A collinear mode only affects the $p_T$-distribution, as the contribution to $\Tau$ from its small light-cone component is power suppressed. Similarly, a soft mode only contributes to $\Tau$, whereas the collinear-soft modes contributes to both measurements. These on-shell modes are uniquely specified by these features. Of course, additional (redundant) modes may be included, as long as the double counting is removed (for example by an appropriate zero-bin subtraction~\cite{Manohar:2006nz}).
As usual, we will assume the cancellation of (off-shell) Glauber modes. These account for initial-state hadron-hadron interactions taking place before the collision, which would ruin factorization~\cite{Bodwin:1981fv}. This cancellation has only been rigorously proven for inclusive Drell-Yan~\cite{Collins:1988ig}, and could be spoiled due to our $p_T$ and $\Tau$ measurements~\cite{Gaunt:2014ska}. 
 
The QCD quark and gluon fields are decomposed into several SCET fields which scale differently with respect to the expansion parameter $\lambda$. By matching quark currents from QCD onto \SCETab we obtain
   \begin{align}  \label{eq:matching}
      \bar \Psi \,\Gamma\, \Psi = C(Q^2,\mu)\,  \bar \xi_\bn W_\bn  S_\bn^\dagger X_\bn^\dagger V_\bn\, \Gamma\, V_n^\dagger X_n S_n W_n^\dagger \xi_n
   \,.\end{align}
The matching coefficient $C(Q^2,\mu)$  captures the effect of hard virtual gluon exchanges not present in the effective theory.
  In \eq{matching}, $\xi_n$ and $\bar \xi_\bn$ are the fields for collinear (anti-)quarks moving in the $n$ ($\bn$) direction and $\Ga$ denotes a generic Dirac structure.
   The Wilson line $W_n$ arises from $n$-collinear gluons emitted by $\bar \Psi$ (which itself is $\bn$-collinear)~\cite{Bauer:2001ct}
\begin{align} 
W_n &=  P\exp\biggl[\,\img g\int_{-\infty}^0 \df u\, \bn\sdt A_n(u\,\bn) \biggr]  
\,.
\end{align}
 The Wilson line $V_n$ is its direct analog for $n$-collinear-soft gluons (obtained by replacing $A_n \to A_{ncs}$). Soft gluons emitted by $\Psi$ are summed into the Wilson line $S_n$~\cite{Bauer:2001yt}
\begin{align} 
S_n &=  P\exp\biggl[\,\img g\int_{-\infty}^0 \df u\, n\sdt A_s(u\,n) \biggr]  
\,,
\end{align}
and the analog for $n$-collinear-soft gluons is $X_n$. 
   
   To fix the ordering of Wilson lines, we exploit gauge invariance of \SCETab. In order to preserve the scaling of the fields, separate collinear, collinear-soft and soft gauge transformations have to be introduced, see e.g.~refs.~\cite{Bauer:2001yt,Bauer:2011uc}.
   Only the $n$-collinear fields transform under $n$-collinear gauge transformations. The other fields are taken far off-shell and are thus unable to resolve the local change induced by this gauge transformation. This causes $W_n^\dagger \xi_n$ and $\bar \xi_\bn W_\bn$ to be grouped together.
Under a $n$-collinear-soft gauge transformation $U_{ncs}$ 
   \begin{align}
     W_n^\dagger \xi_n &\to W_n^\dagger \xi_n
     \,, &  
     S_n &\to S_n 
     \,, &      
     V_n &\to U_{ncs} V_n
     \,, &
     X_n &\to U_{ncs} X_n
     \,, \nn \\
     \bar \xi_\bn W_\bn &\to \bar \xi_\bn W_\bn 
     \,, &  
     S_\bn &\to S_\bn 
     \,, &
     V_\bn &\to V_\bn
     \,, &
     X_\bn &\to X_\bn
   \,,\end{align}
   which groups $V_n^\dagger X_n$ together. Similarly, $X_\bn^\dagger V_\bn$ must be grouped together by $\bn$-collinear-soft gauge invariance.
   The effect of a soft gauge transformation $U_{s}$ is given by
   \begin{align} \label{eq:S_transf}
     W_n^\dagger \xi_n &\to W_n^\dagger \xi_n   
     \,, &          
     S_n &\to U_{s} S_n 
     \,, &      
     V_n &\to U_{s} V_n U_{s}^{\dagger}
     \,, &
     X_n &\to U_{s} X_n U_{s}^{\dagger}
     \,, \nn \\
     \bar \xi_\bn W_\bn &\to \bar \xi_\bn W_\bn 
     \,, &     
     S_\bn &\to U_{s} S_\bn 
     \,, &
     V_\bn &\to U_{s} V_\bn U_{s}^{\dagger}
     \,, &
     X_\bn &\to U_{s}  X_\bn U_{s}^{\dagger}
   \,.\end{align}
The soft gluon field acts as smooth background for collinear-soft modes, implying that the effect of a soft gauge transformation on collinear-soft modes is similar to a global color rotation. 
This almost fixes the ordering in \eq{matching}. There are still a few other possibilities that satisfy the constraints from gauge invariance, such as $\bar \xi_\bn W_\bn  S_\bn^\dagger V_n^\dagger X_n \, \Gamma \, X_\bn^\dagger V_\bn S_n W_n^\dagger \xi_n$. However, these can be ruled out by considering the tree-level matching of QCD onto \SCETab.

   At this point the soft fields still interact with the collinear-soft fields, as indicated in the middle panel of \fig{modes}.
   By performing the analog of the BPS field redefinition~\cite{Bauer:2001yt}, we decouple the soft fields from the collinear-soft fields, 
   \begin{align} \label{eq:BPS}
     V_n &\to S_n V_n S_n^\dagger 
     \,, &     
     X_n &\to S_n X_n S_n^\dagger 
      \,, \nn \\
     V_\bn &\to S_\bn V_\bn S_\bn^\dagger 
     \,, &      
     X_\bn &\to S_\bn X_\bn S_\bn^\dagger 
   \,.\end{align}
   This leads to
   \begin{align} 
      \bar \Psi \,\Gamma \,\Psi = C(Q^2,\mu)\,  \bar \xi_\bn W_\bn  X_\bn^\dagger V_\bn  S_\bn^\dagger \,\Gamma \, S_n V_n^\dagger X_n W_n^\dagger \xi_n
 \,.\end{align}
     The various modes in this matching equation no longer interact and the derivation of factorization formulae now follows the standard procedure in SCET. In particular, establishing factorization to all orders in $\alpha_s$ requires decoupling of the different modes in the Lagrangian, for which we refer to ref.~\cite{Bauer:2011uc}.

One expects that this matching receives power corrections of the size $\la^{2r-1} \sim p_T^2/(Q \Tau)$ and $\la^{2-2r} \sim \Tau^2/p_T^2$, which measure the distance from the respective \SCETa and \SCETb boundary regions of phase space. In our NLO calculation in \sec{NLO} we find corrections of the first type but not of the second. However, we expect that this will no longer be the case at higher orders.
  
  Finally, we briefly comment on the (dis)similarities of our theory with the SCET${}_+$ introduced in ref.~\cite{Bauer:2011uc}. In that paper the dijet invariant mass ($m_{j_1j_2}$) distribution for nearby jets is calculated, with the hierarchy $m_{j_1},m_{j_2} \ll m_{j_1j_2} \ll Q$. Their collinear-soft modes can resolve the two nearby jets, whereas the soft modes do not, and the collinear modes are restricted to the individual jets. Their factorization theorem involves convolutions through the small collinear light-cone component. Since we consider different type of observables, our convolutions of collinear-soft modes with either collinear or soft radiation have a different structure.
The matching in ref.~\cite{Bauer:2011uc} was (also) performed in two steps, where in the first step the two nearby jets are not resolved from each other. Nevertheless, the similarities between the modes and Wilson lines in our and their approach seemed sufficient to us to adopt the same name for our effective theory.
 
\subsection{Factorization Formulae}
\label{sec:fact_formula}

We now discuss SCET factorization formulae for Drell-Yan cross sections that are differential both in $\Tau$ and $p_T$, both at the \SCETa and \SCETb phase space boundaries and in the \SCETab ``bulk".  
In Drell-Yan  production, $pp \to Z/\ga^* \to \ell^+\ell^-$, the lepton pair has a large invariant mass $Q$. A proof of factorization at leading power in $\lqcd/Q$ has been given by Collins, Soper and Sterman~\cite{Collins:1984kg}, for any value of the transverse momentum $p_T$ of the lepton pair, namely for both $p_T \sim Q$ and $p_T  \ll Q$. Here we impose in addition a veto on hard central jets through a cut on beam thrust $\Tau$ in the center-of-mass frame of the $pp$ collision~\cite{Berger:2010xi}, see \eq{Tau}. 
We consider different kinematic regimes for $p_T$ and $\Tau$, as discussed in the introduction. We will not perform the joint resummation of threshold logarithms that becomes important as $Q$ approaches the total CM energy $\Ecm$~\cite{Laenen:2000ij}.

If $\lqcd \ll p_T \sim (\Tau Q)^{1/2} \ll Q$ (\SCETa case), we have the following leading-power factorization formula~\cite{Stewart:2009yx,Jain:2011iu} 
\begin{align} \label{eq:DY_fact_SCETone}
\frac{\df^4 \si}{\df Q^2\, \df Y\, \df p_T^{\,2}\, \df \Tau}
&=  \sum_{q}\,\hat \sigma_{q}^0 \, H(Q^2, \mu)  \!\int\!\df t_1\, \df t_2  \int\! \df^{2} \vec k_{1\perp}\, \df^{2} \vec k_{2\perp}\int\! \df k^+\,S(k^+, \mu) \nn\\ &\quad \times
\Big[ B_q(t_1, x_1, \vec k_{1\perp}, \mu)\, B_{\bar q}(t_2, x_2, \vec k_{2\perp},\mu) + (q\leftrightarrow \bar q) \,\Big] \,
 \nn\\ &\quad \times
\de\Bigl(\Tau - \frac{e^{-Y} t_1 + e^Y t_2}{Q}-k^+\Bigr)\,\de\big(p_T^{\,2} - \abs{\vec k_{1\perp}+\! \vec k_{2\perp}}^{2}\big)
\,,\end{align}
whose ingredients we will describe below.
The sum extends over the various quark flavors, $Y$ is the total rapidity of the leptons, and the momentum fractions of the colliding partons are
\begin{equation}
x_1 = \frac{Q}{\Ecm}\,e^{Y}
\,,\qquad
x_2 = \frac{Q}{\Ecm}\,e^{-Y}
\,.\end{equation}
The quantities $e^{-Y} t_1/Q$, $e^Y t_2/Q$ and $k^+$ in \eq{DY_fact_SCETone} are the contributions to $\Tau$  from the $n$-collinear, $\bn$-collinear and soft radiation. For $n$-collinear radiation, we always have $p_i^+ < p_i^-$, for $\bar n$-collinear radiation, $p_i^+ > p_i^-$, whereas the soft radiation can go into both hemispheres ($p_i^+ < p_i^-$ and $p_i^+ > p_i^-$).

At leading order in the electroweak interactions, 
\begin{equation} \label{eq:si0}
\hat \sigma_{q}^0
= \frac{4\pi \al_\text{em}^2}{9Q^2 \Ecm^2}\bigg[Q_q^2 + \frac{(v_q^2 + a_q^2) (v_\ell^2+a_\ell^2) - 2 Q_q v_q v_\ell (1-m_Z^2/Q^2)}
{(1-m_Z^2/Q^2)^2 + m_Z^2 \Gamma_Z^2/Q^4}\bigg]
\,,\end{equation}
where $Q_q$ is the quark charge in units of $\abs{e}$, $v_{\ell,q}$ and $a_{\ell,q}$ are the standard vector and axial couplings of the leptons and quarks, and $m_Z$ and $\Gamma_Z$ are the mass and width of the $Z$ boson. 

The hard function $H(Q^2)$ is the square of the Wilson coefficient $C(Q^2)$ for the matching of QCD onto SCET vector and axial quark currents\footnote{As compared to~\eq{matching}, in \SCETa only collinear and (ultra-)soft Wilson lines enter the matching.}
\begin{align}
H(Q^2, \mu) =  \abs{C(Q^2, \mu)}^2
\,.\end{align}
It does not depend on $p_T$, since we only consider $p_T \ll Q$.\footnote{The leptonic tensor in the Drell-Yan process does not depend on $p_T$ at leading order.} 
Since lepton masses are neglected, there is no contribution from gluon operators in the matching of the (axial) currents~\cite{Stewart:2009yx}. The gluon PDF only appears through its contribution to the quark beam function, see \eq{FU_B}. 

Due to the \SCETa hierarchy of scales, the effect of soft radiation on the $p_T$-distribution is power suppressed, so only the fully-unintegrated (FU) PDFs account for the recoil of the energetic initial-state radiation against the final-state leptons. Because we consider perturbative $p_T,\Tau \gg \lqcd$, we will refer to these as FU beam functions in the following.
At the bare level, these are defined as the following proton matrix element of collinear fields~\cite{Jain:2011iu} 
\begin{align} \label{eq:B_def}
B_q(t,x,\vec k_\perp)
= 
\MAe{p_n (p^-)}{\bar\chi_n(0)\, \frac{\bnslash}{2} 
  \big[\delta(k^- - p^- + {\bf P}^-)\, \de(t - k^- {\bf P}^+)\, \de^2(\vec k_\perp -  {\vec {\bf P}}_\perp)  \, \chi_n(0)\big]\,}{p_n(p^-)}
\, .
\end{align}
The light-like vector $n^\mu$ is along the direction of the incoming proton (i.e.~$p^\mu = \Ecm n^\mu/2$) and the operator $\bf P$ returns the momentum of the intermediate state.\footnote{We can avoid using the label-momentum formalism employed in e.g.~refs.~\cite{Stewart:2010qs,Jain:2011iu} since after factorization the collinear sector is simply a boosted copy of QCD.} By boost invariance along the $n$-direction, these functions only depend on the momentum fraction $x= k^-/p^-$, the transverse virtuality $-t = k^-k^+$ of the colliding parton, and the transverse momentum $\vec{k}_\perp$~\cite{Stewart:2010qs,Jain:2011iu}.

The (ultra-)soft radiation is described by the beam thrust soft function $S(k)$~\cite{Stewart:2009yx}. This is given in terms of a soft Wilson-line correlator as
\begin{align} 
S(k^+) =
\frac{1}{N_c}\,\langle 0 | {\rm Tr}\big[ {\bf{\overline{T}}} (S_{n}^\dagger(0) S_{\bn}(0))\, \de(k^+ - {\bf P}_1^+ - {\bf P}_2^-)\, {\bf{T}} (S_{\bn}^\dagger(0) S_{n}(0))\big]| 0 \rangle 
\,,\end{align}
where  $({\bf{\overline T}})$ ${\bf T}$ denotes (anti)time ordering and the operator ${\bf P}_1$ (${\bf P}_2$) gives the momentum of the soft radiation going into the hemisphere defined by $p_i^+ < p_i^-$ ($p_i^+ >p_i^-$).

In the region of phase space described by \SCETab ($\lqcd \ll \Tau \ll p_T \ll (\Tau Q)^{1/2} \ll Q$),
\begin{align} \label{eq:DY_fact_SCETonetwo}
\frac{\df^4 \si}{\df Q^2\, \df Y\, \df p_T^{\,2}\, \df \Tau}
&=  \sum_{q}\,\hat \sigma_{q}^0 \, H(Q^2, \mu) \int\! \df^{2} \vec k_{1\perp}\, \df^{2} \vec k_{2\perp}\, \df^{2} \vec k_{1\perp}^{\rm cs}\, \df^{2} \vec k_{2\perp}^{\rm cs}\int\! \df k_1^+\, \df k_2^+\, \df k^+\,S(k^+, \mu) 
\nn\\ &\quad \times
B_q(x_1, \vec k_{1\perp}, \mu, \nu)\, B_{\bar q}(x_2, \vec k_{2\perp},\mu, \nu)\,  \CS\bigl(k_1^+,\vec k_{1\perp}^{\rm cs},\mu,\nu \bigr)\, \CS \bigl(k_2^+,\vec k_{2\perp}^{\rm cs},\mu,\nu \bigr)
\nn \\ & \quad \times
\de\big(\Tau - k_1^+ - k_2^+ - k^+\big)\,\de\big(p_T^{\,2} - \abs{\vec k_{1\perp}+\! \vec k_{2\perp}+\vec k_{1\perp}^{\rm cs}+\! \vec k_{2\perp}^{\rm cs}}^{2}\big)  \!+\! (q\leftrightarrow \bar q)
\,.\end{align}
The contribution from collinear radiation is now encoded in TMD beam functions,
\begin{align} 
B_q(x,\vec k_\perp)
= 
\MAe{p_n (p^-)}{\bar\chi_n(0)\, \frac{\bnslash}{2} 
  \big[\delta(k^- - p^- + {\bf P}^-)\, \de^2(\vec k_\perp -  {\vec {\bf P}}_\perp)  \, \chi_n(0)\big]\,}{p_n(p^-)}
\, \end{align}
Their naive definition using dimensional regularization is known to suffer from light-cone singularities (rapidity divergences), which we regulate following refs.~\cite{Chiu:2011qc,Chiu:2012ir}. There are separate but identical collinear-soft functions for the $n$ and $\bn$ direction,
\begin{align} \label{eq:CSdef}
\CS(k^+,\vec k_\perp) &=
\frac{1}{N_c}\,\langle 0 | {\rm Tr}\big[ {\bf{\overline{T}}} (X_n^\dagger(0) V_n(0))\, \de(k^+ - {\bf P}^+)\,\de^2(\vec k_\perp - \vec {\bf P}_\perp) {\bf{T}} (V_n^\dagger(0) X_{n}(0))\big]| 0 \rangle 
\,, \nn \\
&= \frac{1}{N_c}\,\langle 0 | {\rm Tr}\big[ {\bf{\overline{T}}} (V_\bn^\dagger(0) X_\bn(0))\, \de(k^+ - {\bf P}^-)\,\de^2(\vec k_\perp - \vec {\bf P}_\perp)\, {\bf{T}} (X_\bn^\dagger(0) V_\bn(0))\big]| 0 \rangle 
\,,\end{align}
which are also affected by rapidity divergences.

For the hierarchy $\lqcd \ll p_T \sim \Tau \ll Q$, soft modes have the same virtuality and transverse momentum as the collinear ones, and contribute both to $\Tau$ and $p_T$ measurements. The corresponding  $\SCETb$ factorization theorem has the form
\begin{align} \label{eq:DY_fact_SCETtwo}
\frac{\df^4 \si}{\df Q^2\, \df Y\, \df p_T^{\,2}\, \df \Tau}
&=  \sum_{q}\,\hat \sigma_{q}^0 \, H(Q^2, \mu)  \! \int\! \df^{2} \vec k_{1\perp}\, \df^{2} \vec k_{2\perp}\, \df^{2} \vec k_{\perp}\int\! \df k^+\; \de\big(p_T^{\,2} \!-\! \abs{\vec k_{1\perp} \!+\! \vec k_{2\perp} \!+\!  \vec k_{\perp}}^{2}\big) \de\big(\Tau \!-\! k^+ \big)
\nn\\ &\quad \times
\Big[ B_q(x_1, \vec k_{1\perp}, \mu, \nu)\, B_{\bar q}(x_2, \vec k_{2\perp},\mu, \nu) + (q\leftrightarrow \bar q) \,\Big]\,S(k^+, \vec{k}_{\perp}, \mu, \nu)
\,.\end{align}
The new ingredient is given by the FU soft function, which is defined as
\begin{equation} \label{eq:S_def}
S(k^+, \vec k_\perp) =
\frac{1}{N_c}\,\langle 0 | {\rm Tr}\big[ {\bf{\overline{T}}} (S_{n}^\dagger(0) S_{\bn}(0))\, \de(k^+ - {\bf P}_1^+ - {\bf P}_2^-)\,\de^2(\vec k_\perp - \vec {\bf P}_\perp)\, {\bf{T}} (S_{\bn}^\dagger(0) S_{n}(0))\big]| 0 \rangle 
\,.\end{equation}

It is natural to ask to what extent our approach can be used to calculate non-global logarithms, which arise when different restrictions are applied to distinct regions of phase space~\cite{Dasgupta:2001sh,Dasgupta:2002dc}. If instead of the transverse momentum of the $Z$ boson one measures the $p_{T,{\rm ISR}}$ of the initial-state radiation that recoils against it, we could e.g.~restrict ourselves to the ISR in \emph{one} hemisphere. In this case the factorization theorem in the region of phase space described by \SCETab is simply modified to
\begin{align} \label{eq:DY_fact_SCETonetwo_NGL}
\frac{\df^4 \si}{\df Q^2\, \df Y\, \df p_{T,{\rm ISR}}^{\,2}\, \df \Tau}
&=  \sum_{q}\,\hat \sigma_{q}^0 \, H(Q^2, \mu) \int\! \df t_2\, \int\! \df^{2} \vec k_{1\perp}\, \df^{2} \vec k_{1\perp}^{\rm cs} \int\! \df k_1^+\, \df k^+\,S(k^+, \mu) 
\nn\\ &\quad \times
B_q(x_1, \vec k_{1\perp}, \mu, \nu)\, B_{\bar q}(t_2,x_2,\mu)\,  \CS\bigl(k_1^+,\vec k_{1\perp}^{\rm cs},\mu,\nu \bigr)
\nn \\ & \quad \times
\de\big(\Tau - k_1^+ - \frac{e^Y t_2}{Q} - k^+\big)\,\de\big(p_{T,{\rm ISR}}^{\,2} - \abs{\vec k_{1\perp}+\vec k_{1\perp}^{\rm cs}}^{2}\big)  \!+\! (q\leftrightarrow \bar q)
\,.\end{align}
However, this does not address the problem arising when the \emph{soft function} contains multiple scales (see for example~\cite{Kelley:2011ng,Hornig:2011iu,Hornig:2011tg}), which occurs when e.g.~the beam thrust measurement is restricted to one hemisphere.

\section{Ingredients at NNLL}
\label{sec:ingredients}

In this section we collect the expressions for the ingredients entering the factorization formulae in \sec{fact_formula}, to the accuracy needed for NNLL resummations: the hard function at one loop is discussed in \sec{hard}, the FU and TMD beam function in \sec{beam}, the FU and beam thrust soft function in \sec{soft} and the collinear-soft function in \sec{cs}. The FU soft function and collinear-soft function are calculated for the first time. RG equations and anomalous dimensions for NNLL resummation are given in \sec{renormalization} and \app{RGE}. The anomalous dimensions of the collinear-soft function and FU soft function satisfy the consistency requirement imposed by the $\mu$ and $\nu$ independence of the factorized cross sections in \eqs{DY_fact_SCETonetwo}{DY_fact_SCETtwo}. In \sec{nll} we combine these ingredients to obtain a compact expression for the NLL cross section.

\subsection{Hard Function}
\label{sec:hard}

The one-loop Wilson coefficient $C(Q^2,\mu)$ from matching the quark current in QCD onto SCET was computed in refs.~\cite{Manohar:2003vb, Bauer:2003di}. Here $Q^2$ is the square of the partonic center of mass energy. The matching is the same for \SCETa, \SCETab and \SCETb, because all effective field theory diagrams are scaleless and vanish in dimensional regularization. At one loop, 
\begin{equation} \label{eq:H_oneloop}
H(Q^2, \mu) 
= \Abs{C(Q^2, \mu)}^2 = 1 + \frac{\alpha_s C_F}{2\pi} \biggl[-\ln^2 \Bigl(\frac{Q^2}{\mu^2}\Bigr) + 3 \ln \Bigl(\frac{Q^2}{\mu^2}\Bigr) - 8 + \frac{7\pi^2}{6} \biggr]
\,.\end{equation}

\subsection{Beam Functions}
\label{sec:beam}
  
The FU beam function was defined in \eq{B_def}, and its arguments $t$ and $\vks$ are restricted to be of the same parametric size.
  As we assume that these scales are perturbative, the FU beam function can be matched onto PDFs~\cite{Stewart:2009yx,Stewart:2010qs,Mantry:2010mk,Jain:2011iu}
  \begin{align} \label{eq:FU_B}
  B_q(t,x,\vec k_\perp,\mu) 
  & = \sum_{j= u, \bar u, d, g \dots} \int_x^1 \frac{\df x'}{x'}
   \cI_{qj}\Big(t,\frac{x}{x'},\vec k_\perp,\mu\Big) f_j(x',\mu) \bigg[1 + \ORd{\frac{\lqcd^2}{t},\frac{\lqcd^2}{\vks}} \bigg]
\,.  \end{align}
  Because of the kinematic bound $\vks \leq (1-x)t/x$ (see eq.~(1.1) of ref.~\cite{Jain:2011iu}), the renormalization is the same as the standard beam function and
  \begin{align} \label{eq:FU_int}
   \int\! \df^2 \vec k_\perp\, B_q(t,x,\vec k_\perp,\mu) = B_q(t,x,\mu)
  \,.\end{align}
  Up to NLO, the matching coefficients in \eq{FU_B} are~\cite{Jain:2011iu} 
\begin{align} 
\cI_{qq}^\zero(t,x,\vec k_\perp,\mu)
& =  \de(t)\, \de(1-x)\, \de^2(\vec k_\perp) 
\,,\nn \\
\cI_{qg}^\zero(t,x,\vec k_\perp,\mu)
& = 0
\,,\nn \\
\cI_{qq}^\one(t,x,\vec k_\perp,\mu)
& = \frac{\al_s(\mu) C_F}{2\pi^2}\bigg\{
\frac{2}{\mu^2} \cL_1\Big(\frac{t}{\mu^2}\Big)\, \de(1\!-\!x)\, \de(\vks)
+ \frac{1}{\mu^2} \cL_0\Big(\frac{t}{\mu^2}\Big)\, (1\!+\!x^2)\, \cL_0(1\!-\!x)\,
\de\Big(\vks \!-\! \frac{(1\!-\!x) t}{x} \Big) 
\nn \\ & \quad 
+ \de(t)\, \de(\vks) \Big[
(1+x^2) \cL_1(1-x) - \frac{\pi^2}{6}\, \de(1-x) - \frac{1+x^2}{1-x}\, \ln x + 1-x \Big] \bigg\}
\,, \nn \\
\cI_{qg}^\one(t,x,\vec k_\perp,\mu)
& =  \frac{\al_s(\mu) T_F}{2\pi^2}\bigg\{\Big[\frac{1}{\mu^2} \cL_0\Big(\frac{t}{\mu^2}\Big)\,\de\Big(\vks \!-\! \frac{(1\!-\!x)t}{x} \Big) + \de(t)\, \de(\vks) \ln \frac{1-x}{x} \Big]\big[x^2\!+\!(1\!-\!x)^2\big] 
 \nn \\ & \quad
+ 2\,\de(t)\, \de(\vks) \,x(1-x)  \bigg\} 
\,,\end{align}
where some additional factors of $1/\pi$ are due to
\begin{align}
\de^2(\vec k_\perp) = \frac{1}{\pi}\, \de(\vks)
\,.\end{align}
The matching coefficients at NNLO have recently been calculated in ref.~\cite{Gaunt:2014xxa}.

  The TMD beam function satisfies a similar equation~\cite{Collins:1981uw,Collins:1984kg,Becher:2010tm,Chiu:2012ir}
  \begin{align}
  B_q(x,\vec k_\perp,\mu,\nu) 
  & = \sum_j \int_x^1 \frac{\df x'}{x'}
   \cI_{qj}\Big(\frac{x}{x'},\vec k_\perp,\mu,\nu \Big) f_j(x',\mu) \bigg[1+\ORd{\frac{\lqcd^2}{\vks}}\bigg]
\,,  \end{align}
with coefficients~\cite{Ritzmann:2014mka}
\begin{align}
\cI_{qq}^\zero(x,\vec k_\perp,\mu,\nu)
& = \de(1-x)\, \de^2(\vec k_\perp) 
\,,\nn \\
\cI_{qg}^\zero(x,\vec k_\perp,\mu,\nu)
& = 0
\,,\nn \\
\cI_{qq}^\one(x,\vec k_\perp,\mu,\nu) &= 
\frac{\al_s C_F}{2\pi^2} \bigg\{ \frac{1}{\mu^2} \cL_0\Big(\frac{\vec k_\perp^{\,2}}{\mu^2}\Big) \bigg[(1+x^2) \cL_0(1-x) 
+ 2\, \de(1-x) \ln \frac{p^-}{\nu} \bigg] +
\de(\vec k_\perp^{\,2}) (1-x) \bigg\}
\,,\nn \\
\cI_{qg}^\one(x,\vec k_\perp,\mu,\nu) &= 
\frac{\al_s T_F}{2\pi^2} \bigg\{ \frac{1}{\mu^2} \cL_0\Big(\frac{\vec k_\perp^{\,2}}{\mu^2}\Big) \big[x^2+(1-x)^2\big]
+ 2\,\de(\vec k_\perp^{\,2}) x(1-x) \bigg\}
\,.\end{align}
Most approaches (such as in refs.~\cite{Becher:2010tm,Collins:2011zzd,GarciaEchevarria:2011rb}) do not (need to) separate the TMD beam and TMD soft function. In the \SCETab regime, instead, we need the TMD beam function but have a different soft function. 

\subsection{Soft Functions}
\label{sec:soft}

The (beam) thrust soft function was determined at NLO in refs.~\cite{Schwartz:2007ib, Fleming:2007xt,Stewart:2009yx}
\begin{align}      
S(k^+,\mu) &= \de(k^+) + \frac{\alpha_s\,C_F}{2\pi} \biggl[
-\frac{8}{\mu} \cL_1\Big(\frac{k^+}{\mu}\Big) +
\frac{\pi^2}{6}\, \delta(k^+) \biggr] + \ord{\al_s^2}
\,.\end{align}
The NNLO contribution is known as well~\cite{Hornig:2011iu,Kelley:2011ng}.

We now calculate the FU soft function, which is differential in both $k^+$ and $\vec k_\perp$, with $k^+ \sim | \vec k_\perp|$.\footnote{This differs from the FU soft function in ref.~\cite{Mantry:2009qz}, because their $k^+$ measurement is independent of the hemisphere the gluon goes into.}
Starting from the definition in \eq{S_def}, the tree-level result is
\begin{align}
  S^\zero(k^+,\vec k_\perp) = \de(k^+)\,\de^2(\vec k_\perp)
\,.\end{align}
Using the rapidity regulator of refs.~\cite{Chiu:2011qc,Chiu:2012ir}, at one-loop order we find
      \begin{align} \label{eq:FUS_bare}
        S^\one(k^+,\vec k_\perp) &= \frac{4 g^2 w^2 C_F}{(2\pi)^{3-2\eps}} \Big(\frac{e^{\ga_E} \mu^2}{4\pi}\Big)^\eps\, \nu^\eta
        \int\! \df^d \ell\, \theta(\ell^0) \de(\ell^2)\, \frac{|2 \ell^3|^{-\eta}}{\ell^- \ell^+}
        \nn \\ & \quad \times
        \de^2(\vec \ell_\perp - \vec k_\perp) \,
        \de \big(\ell^+ \theta(\ell^- - \ell^+) + \ell^- \theta(\ell^+ - \ell^-) - k^+\big)
        \nn \\
       &=\al_s w^2 C_F\, \frac{2^{1-\eta}\, e^{\eps \ga_E}\, \mu^{2\eps}\, \nu^\eta}{\pi^{2-\eps}}
        \int\! \df^{-2\eps} \ell_\eps\, \frac{1}{\vec k_\perp^{\,2} + \vec \ell_\eps^{\,2}}
        \nn \\ & \quad \times        
        \int_0^\infty\! \df \ell^3\, \frac{(\ell^3)^{-\eta}}{\sqrt{\vec k_\perp^{\,2} + \vec \ell_\eps^{\,2} + (\ell^3)^2}}\,
        \de \Big(\sqrt{\vec k_\perp^{\,2} + \vec \ell_\eps^{\,2} + (\ell^3)^2} - \ell^3 - k^+\Big)
        \nn \\
       &=\al_s w^2 C_F\, \frac{2\, e^{\eps \ga_E}\, \mu^{2\eps}\, \nu^\eta}{\pi^{2-\eps}}\, \frac{1}{(k^+)^{1-\eta}}
        \int\! \df^{-2\eps} \ell_\eps\, \frac{\theta\big(\vec k_\perp^{\,2} - (k^+)^2 + \vec \ell_\eps^{\,2} \big)}{(\vec k_\perp^{\,2} + \vec \ell_\eps^{\,2})[\vec k_\perp^{\,2} - (k^+)^2 + \vec \ell_\eps^{\,2}]^\eta}
        \nn \\
       &=\al_s w^2 C_F\, \frac{2\, e^{\eps \ga_E}\, \mu^{2\eps}\, \nu^\eta}{\pi^2\, \Ga(-\eps)}\, \frac{1}{(k^+)^{1-\eta} (\vks)^{1+\eps+\eta}}
       \int_{0}^\infty\!\! \df x\, \frac{\theta(x+1-(k^+)^2/\vks)}{x^{1+\eps}(x + 1)[x + 1 - (k^+)^2/\vks]^\eta}
       \nn \\ 
       &=\al_s w^2 C_F\, \frac{2\, e^{\eps \ga_E}\, \mu^{2\eps+\eta}\, \nu^\eta}{\pi^2\, \Ga(-\eps)}\, 
       \bigg\{\frac{1}{\eta} \de(k^+) \, \frac{1}{(\vks)^{1+\eps+\eta}}
       \int_{0}^\infty\!\! \df x\, \frac{1}{x^{1+\eps}(x + 1)^{1+\eta}}
       \nn \\ & \quad
       + \frac{1}{\mu} \cL_0\Big(\frac{k^+}{\mu}\Big) \frac{1}{(\vks)^{1+\eps}}
       \bigg[
       \theta(\vks - (k^+)^2) \int_{0}^\infty\!\! \df x\, \frac{1}{x^{1+\eps}(x + 1)} 
       \nn \\ & \quad
       + \theta((k^+)^2 - \vks) \int_{(k^+)^2/\vks-1}^\infty\!\!\! \df x\, \frac{1}{x(x + 1)}\bigg]+ \ord{\eta,\eps}\bigg\}
      \nn \\
      &\stackrel{\vks>0}{=}\frac{\al_s w^2 C_F}{\pi^2}
       \bigg\{
       \frac{2}{\eta}\bigg[
       - \frac{1}{\eps}\, \delta(\vks) + \frac{1}{\mu^2} \cL_0\Big(\frac{\vks}{\mu^2}\Big) \bigg] \de(k^+)
       + \frac{2}{\eps^2}\, \delta(\vks) \de(k^+)
       \nn \\ & \quad
       +\frac{2}{\eps}\, \ln \frac{\mu}{\nu}  \delta(\vks) \de(k^+)
       +2 \theta(\vks - (k^+)^2)\,\frac{1}{\mu} \cL_0\Big(\frac{k^+}{\mu}\Big) \frac{1}{\mu^2} \cL_0\Big(\frac{\vks}{\mu^2}\Big)
       \nn \\ & \quad
       + \de(k^+) \bigg[
       - \frac{2}{\mu^2} \cL_1\Big(\frac{\vks}{\mu^2}\Big)
       +\frac{2}{\mu^2} \cL_0\Big(\frac{\vks}{\mu^2}\Big) \ln \frac{\nu}{\mu}
       -\frac{\pi^2}{6}  \delta(\vks)\bigg] + \ord{\eta,\eps}\bigg\} 
      \nn \\
      &\stackrel{\vks\geq0}{\to} \frac{\al_s w^2 C_F}{\pi^2}
       \bigg\{
       \frac{2}{\eta}\bigg[
       - \frac{1}{\eps}\, \delta(\vks) + \frac{1}{\mu^2} \cL_0\Big(\frac{\vks}{\mu^2}\Big) \bigg] \de(k^+)
       + \frac{1}{\eps^2}\, \delta(\vks) \de(k^+)
       \nn \\ & \quad
       +\frac{2}{\eps}\, \ln \frac{\mu}{\nu}  \delta(\vks) \de(k^+)
       +\frac{2}{\mu^3} \cL_\Delta\Big(\frac{k^+}{\mu},\frac{\vks}{\mu^2}\Big)
       + \de(k^+) \bigg[
       - \frac{2}{\mu^2} \cL_1\Big(\frac{\vks}{\mu^2}\Big)
       \nn \\ & \quad
       +\frac{2}{\mu^2} \cL_0\Big(\frac{\vks}{\mu^2}\Big) \ln \frac{\nu}{\mu}
       -\frac{\pi^2}{12}  \delta(\vks)\bigg] + \ord{\eta,\eps}\bigg\} 
      \,.\end{align}
Here longitudinal momenta get regulated by $\eta$, which can be thought of as the analog for rapidity divergences of the UV regulator $\eps$, with the dimensionful parameter $\nu$ acting like a renormalization scale. Both $1/\eta$ and $1/\eps$ divergences get absorbed in renormalization constants and give rise to $\mu$- and $\nu$-RG equations. The bookkeeping parameter $w$ is used to derive the anomalous dimensions (see \eq{renorm}) and will be eventually set equal to 1.    

     In \eq{FUS_bare} we introduce $x= \vec \ell_\eps^{\,2} / \vks$  in intermediate steps, to simplify notation. 
     In the second to last step, we first assume $\vks > 0$ to simplify the expansion in $\eps$. We then extend the distributions to include $\vks=0$ and fix the coefficient of the $\delta(k^+)\delta(\vks)$ by integrating the \emph{unexpanded} result.
    The finite terms contain the following two-dimensional plus distribution
      \begin{align}
        \cL_\Delta(x_1,x_2) &= 
        \lim_{\bt \to 0} \frac{\df}{\df x_1}\, \frac{\df}{\df x_2}\, 
        \Big[\theta(x_2 - x_1^2) \theta(x_1-\bt) \ln x_1\, (\ln x_2 - \ln x_1)
        \nn \\ & \quad
        + \frac14\,\theta(x_1^2 - x_2) \theta(x_2-\bt^2) \ln^2 x_2\Big]
      \,.\end{align}
     The $1/\eps$ and $1/\eta$ poles are renormalized. We obtain the one-loop anomalous dimension in \eq{gamma_FUS} by using~\cite{Chiu:2011qc,Chiu:2012ir} 
     \begin{align} \label{eq:renorm}
      \frac{\df \al_s}{\df \ln \mu} &= -2 \eps\, \al_s + \ord{\al_s^2}
      \,, \nn \\ 
      \frac{\df w}{\df \ln \nu} &= - \frac{\eta}{2}\, w + \ord{w^2}
      \,, \end{align}
      and setting $w=1$ afterwards. These are the same as for the TMD soft function. The renormalized FU soft function is given by the remaining finite terms, 
      \begin{align} \label{eq:FUS}
        S^\one(k^+,\vec k_\perp,\mu,\nu) &=
        \frac{\al_s C_F}{\pi^2}
       \bigg\{\frac{2}{\mu^3} \cL_\Delta\Big(\frac{k^+}{\mu},\frac{\vks}{\mu^2}\Big)
       \nn \\ & \quad
       + \de(k^+) \bigg[
       - \frac{2}{\mu^2} \cL_1\Big(\frac{\vks}{\mu^2}\Big)
       +\frac{2}{\mu^2} \cL_0\Big(\frac{\vks}{\mu^2}\Big) \ln \frac{\nu}{\mu}
       -\frac{\pi^2}{12}  \delta(\vks)\bigg] \bigg\} 
      \,.\end{align}
     Its integral over $k^+$ reproduces the TMD soft function in refs.~\cite{Chiu:2012ir,Ritzmann:2014mka}
     \begin{align}
       \int\! \df k^+\, S^\one(k^+,\vec k_\perp,\mu,\nu) &= 
 \frac{\al_s C_F}{\pi^2} \bigg[- \frac{1}{\mu^2} \cL_1\Big(\frac{\vec k_\perp^{\,2}}{\mu^2}\Big) 
+ \frac{1}{\mu^2} \cL_0\Big(\frac{\vec k_\perp^{\,2}}{\mu^2}\Big) \ln \frac{\nu^2}{\mu^2}
 - \frac{\pi^2}{12} \de(\vec k_\perp^{\,2})\bigg]
     \nn \\ &
     = S^\one(\vec k_\perp,\mu,\nu)       
     \,,\end{align}
    which parallels \eq{FU_int} for the FU beam function.
     Here we used that for $x_1^2 > x_2$,
     \begin{align}
      \int_0^{x_1}\!\df x_1'\,\cL_\Delta(x_1', x_2) =   \lim_{\bt \to 0} \frac{\df}{\df x_2}\, 
        \Big[ \frac14\, \theta(x_2-\bt^2) \ln^2 x_2\Big] = \frac12 \cL_1(x_2)
     \,.\end{align}

\subsection{Collinear-Soft Function}
\label{sec:cs}

The calculation of the collinear-soft function, defined in \eq{CSdef}, is actually quite similar to that of the FU soft function. The main difference is that collinear-soft radiation only goes into one hemisphere, leading to the change
\begin{align}
  \de \big(\ell^+ \theta(\ell^- - \ell^+) + \ell^- \theta(\ell^+ - \ell^-) - k^+\big) \to \de(\ell^+ - k^+)
\,.\end{align}
We conveniently separate out a contribution $\frac12 S^\one(k^+,\vec k_\perp)$ from the hemisphere where the measurement in the FU soft function and collinear-soft function are the same. The remainder does not contain rapidity divergences, allowing us to set $\eta=0$ from the beginning,
      \begin{align} \label{eq:CS_bare}
        \CS^\one(k^+,\vec k_\perp) &= \frac{4 g^2 w^2 C_F}{(2\pi)^{3-2\eps}} \Big(\frac{e^{\ga_E} \mu^2}{4\pi}\Big)^\eps\, \nu^\eta
        \int\! \df^d \ell\, \theta(\ell^0) \de(\ell^2)\, \frac{|2 \ell^3|^{-\eta}}{\ell^- \ell^+}\,
        \de^2(\vec \ell_\perp - \vec k_\perp) \,
        \de \big(\ell^+ - k^+\big)
        \nn \\
       &= \frac12\,S^\one+       
       \al_s w^2 C_F\, \frac{e^{\eps \ga_E}\, \mu^{2\eps}}{\pi^{2-\eps}}
        \int\! \df^{-2\eps} \ell_\eps\, \frac{1}{\vec k_\perp^{\,2} + \vec \ell_\eps^{\,2}}   
        \nn \\ & \quad \times
        \int_0^\infty\! \df \ell^3\, \frac{\de \big(\sqrt{\vec k_\perp^{\,2} + \vec \ell_\eps^{\,2} + (\ell^3)^2} + \ell^3 - k^+\big)}{\sqrt{\vec k_\perp^{\,2} + \vec \ell_\eps^{\,2} + (\ell^3)^2}}
        \nn \\
       &=\frac12\,S^\one +  \al_s w^2 C_F\, \frac{e^{\eps \ga_E}\, \mu^{2\eps}}{\pi^{2-\eps}}\, \frac{1}{k^+}
        \int\! \df^{-2\eps} \ell_\eps\, \frac{\theta\big((k^+)^2 - \vec k_\perp^{\,2} - \vec \ell_\eps^{\,2} \big)}{\vec k_\perp^{\,2} + \vec \ell_\eps^{\,2}}
       \nn \\
       &=\frac12\,S^\one + \al_s w^2 C_F\, \frac{e^{\eps \ga_E}\, \mu^{2\eps}}{\pi^2\, \Ga(-\eps)}\, \frac{\theta(k^+- |\vec k_\perp|)}{k^+}\,
        \int_{0}^{(k^+)^2 - \vec k_\perp^{\,2}}\!\! \df \vec \ell_\eps^{\,2}\, \frac{1}{(\vec \ell_\eps^{\,2})^{1+\eps}(\vec k_\perp^{\,2} + \vec \ell_\eps^{\,2})}
       \nn \\
       &=\frac12\,S^\one + \al_s w^2 C_F\, \frac{e^{\eps \ga_E}\, \mu^{2\eps}}{\pi^2\, \Ga(1-\eps)}\, \frac{\theta(k^+- |\vec k_\perp|)}{[(k^+)^2/\vks-1]^\eps k^+\, (\vks)^{1+\eps}} 
       \nn \\ & \quad \times
        {}_2F_1\Big(1,-\eps,1-\eps,1-\frac{(k^+)^2}{\vks}\Big)   
       \nn \\  
       &=\frac12\,S^\one + \frac{\al_s w^2 C_F}{\pi^2}\,  
       \bigg\{\frac{1}{2\eps^2} \de(k^+) \de(\vks)  - \frac{1}{\eps}\, \frac{1}{\mu} \cL_0\Big(\frac{k^+}{\mu}\Big)\,\de(\vks)
       + \frac{1}{\mu^3}\,\cL_\nabla\Big(\frac{k^+}{\mu},\frac{\vks}{\mu^2}\Big)
       \nn \\ & \quad
   + \de(k^+ - |\vec k_\perp|) \bigg[ \frac{2}{\mu} \cL_1\Big(\frac{k^+}{\mu}\Big) - \frac12\, \frac{1}{\mu^2} \cL_1\Big(\frac{\vks}{\mu^2}\Big) \bigg]
        -\frac{\pi^2}{12}  \de(k^+) \delta(\vks)
 + \ord{\eps}  \bigg\}
 \nn \\
   &=  \frac{\al_s w^2 C_F}{\pi^2}\bigg\{
       \frac{1}{\eta}\bigg[
       - \frac{1}{\eps}\, \delta(\vks) + \frac{1}{\mu^2} \cL_0\Big(\frac{\vks}{\mu^2}\Big) \bigg] \de(k^+)
       + \frac{1}{\eps^2}\, \de(k^+) \delta(\vks) 
       \nn \\ & \quad
       - \frac{1}{\eps}\, \frac{1}{\mu} \cL_0\Big(\frac{k^+}{\mu}\Big)\,\de(\vks)       
       +\frac{1}{\eps}\, \ln \frac{\mu}{\nu}  \delta(\vks) \de(k^+)
       +\frac{1}{\mu} \cL_0 \Big(\frac{k^+}{\mu}\Big) \frac{1}{\mu^2} \cL_0 \Big(\frac{\vks}{\mu^2}\Big)
       \nn \\ & \quad
       + \de(k^+) \bigg[
       - \frac{1}{\mu^2} \cL_1\Big(\frac{\vks}{\mu^2}\Big)
       +\frac{1}{\mu^2} \cL_0\Big(\frac{\vks}{\mu^2}\Big) \ln \frac{\nu}{\mu}
       -\frac{\pi^2}{12}  \delta(\vks)\bigg] + \ord{\eta,\eps}\bigg\} 
\,.\end{align}
      The expansion in $\eps$ is again subtle at $(k^+, \vks)=(0,0)$. Similar to \sec{soft},
      we first expand assuming $k^+ > 0$ and then extend the plus distributions to $k^+=0$, fixing the coefficient of $\de(k^+) \delta(\vks)$ by integration.
      In an intermediate expression, the following two-dimensional plus distribution arises
      \begin{align}
        \cL_\nabla(x_1,x_2) =
        \lim_{\bt \to 0} \frac{\df}{\df x_1}\, \frac{\df}{\df x_2}\, 
        \Big[&\theta(x_1^2 - x_2) \theta(x_2-\bt^2) \Big(\ln x_1 - \frac14 \ln x_2\Big) \ln x_2
        \nn \\ 
        &+ \theta(x_2 - x_1^2) \theta(x_1-\bt) \ln^2 x_1 \Big]
      \,.\end{align}
       In the final expression this combines with $\cL_\Delta$ in \eq{FUS_bare} to give
       \begin{align}
         \cL_\Delta(x_1,x_2) + \cL_\nabla(x_1,x_2)
         = \cL_0(x_1) \cL_0(x_2)
       \,.\end{align}
 The divergences in \eq{CS_bare} lead to the one-loop anomalous dimensions in \eq{gamma_CS}. This satisfies the relation among anomalous dimensions required by consistency of the factorization theorem in \eq{DY_fact_SCETonetwo} at this order. The finite terms give
 \begin{align}
  \CS^\one(k^+,\vec k_\perp,\mu,\nu)  &=  \frac{\al_s C_F}{\pi^2}\bigg\{
       \frac{1}{\mu} \cL_0 \Big(\frac{k^+}{\mu}\Big) \frac{1}{\mu^2} \cL_0 \Big(\frac{\vks}{\mu^2}\Big)
       \nn \\ & \quad
       + \de(k^+) \bigg[
       - \frac{1}{\mu^2} \cL_1\Big(\frac{\vks}{\mu^2}\Big)
       +\frac{1}{\mu^2} \cL_0\Big(\frac{\vks}{\mu^2}\Big) \ln \frac{\nu}{\mu}
       -\frac{\pi^2}{12}  \delta(\vks)\bigg] \bigg\}  
\,. \end{align}

\subsection{Renormalization and Anomalous Dimensions}
\label{sec:renormalization}

In this section we write down the RG equations (RGEs) for all these ingredients, which are well-known except for the FU soft function and collinear-soft function. Their anomalous dimensions are constrained by consistency of the factorization theorems in \sec{fact_formula} and agree with the one-loop calculations in \secs{soft}{cs}.
For completeness we give the expressions for both the quark and gluon case, as indicated by an additional index $i=q,g$ in this section. The anomalous dimensions involve the cusp anomalous dimension $\Ga_\text{cusp}^i$ and non-cusp anomalous dimensions $\ga_H^i, \ga_J^i, \ga_\nu^i$, which are tabulated in \app{RGE}.

The anomalous dimension of the Wilson coefficient $C$ is 
  \begin{align} \label{eq:C_RGE}
\mu \frac{\df}{\df\mu}\, C(Q^2, \mu) &= \gamma_H(Q^2, \mu)\, C(Q^2, \mu) 
\,, \nn \\
\gamma_H(Q^2, \mu) &=
\Gamma_\cusp^q(\al_s) \ln\frac{- Q^2 - \img 0}{\mu^2} + \gamma_H^q(\al_s)
\,, 
\end{align}
from which the evolution of the hard function $H(Q^2,\mu) = |C(Q^2,\mu)|^2$ directly follows.

The FU beam function satisfies the following RGE\footnote{
The additional spin structure~\cite{Mantry:2009qz} for the gluon beam function does not mix under renormalization and satisfies the same RGE.}
\begin{align}
\mu \frac{\df}{\df \mu}\, B_i(t, x, \vec k_\perp,\mu) &= \int_0^t\! \df t'\, \gamma_B^i(t-t',\mu)\, B_i(t', x, \vec k_\perp,\mu)
\,,\nn\\
\gamma_B^i(t, \mu)
&= -2 \Gamma^i_{\cusp}(\alpha_s)\,\frac{1}{\mu^2}\cL_0\Bigl(\frac{t}{\mu^2}\Bigr) + \gamma_J^i(\alpha_s)\,\delta(t)
\,.\end{align}
The TMD beam function also involves a $\nu$ evolution (rapidity resummation)\footnote{Its non-cusp $\mu$-anomalous dimension has not yet been calculated at two loops and is not fixed by consistency. However, the remaining degeneracy is irrelevant, since the TMD beam function has the same $\mu$ scale as the collinear-soft function (in SCET${}_+$) or FU soft function (in SCET${}_{\rm II}$).}
\begin{align} \label{eq:ga_TMD_B}
\mu \frac{\df}{\df \mu}\, B_i(x, \vec k_\perp,\mu,\nu) &= \gamma_B^i(p^-, \mu, \nu)\, B_i(x, \vec k_\perp, \mu, \nu)
\,,\nn\\
\nu \frac{\df}{\df \nu}\, B_i(x, \vec k_\perp,\mu,\nu) &= \int\! \df^2 \vec k_\perp'\, \gamma_\nu^i(\vec k_\perp - \vec k_\perp',\mu)\, B_i(x, \vec k_\perp',\mu,\nu)
\,,\nn\\
\gamma^i_B(p^-, \mu, \nu)
&= 2 \Gamma_\cusp^i(\al_s)\, \ln \frac{\nu}{p^-} + \ga_J^i(\al_s) 
\,, \nn \\
\gamma^i_\nu(\vec k_\perp, \mu)   
&= -  \Gamma_\cusp^i(\al_s)\, \frac{1}{\pi}\,\frac{1}{\mu^2} \cL_0\Big(\frac{\vks}{\mu^2}\Big)
 + \ga_\nu^i(\al_s)\,\de^2 (\vec k_\perp)\,. 
\end{align}

The RGE of the (beam) thrust soft function is given by
\begin{align} 
\mu\frac{\df}{\df\mu}\, S_i(k^+, \mu)
&= \int_0^{k^+}\! \df {k'}^+\, \gamma_S^i(k^+\! - {k'}^+, \mu)\, S_i({k'}^+, \mu)
\,, \nn \\
\gamma_S^i(k^+, \mu)
&= 4\Gamma_\cusp^i(\alpha_s)\, \frac{1}{\mu} \cL_0\Big(\frac{k^+}{\mu}\Big) 
-2\big[\gamma_H^i(\alpha_s) + \gamma_J^i(\alpha_s)\big]\, \delta(k^+)
\,,\end{align}
and for the FU soft function it is given by,
\begin{align} \label{eq:gamma_FUS}
\mu \frac{\df}{\df \mu}\, S_i(k^+,\vec k_\perp,\mu,\nu) &= \gamma_S^i(\mu, \nu)\, S_i(k^+, \vec k_\perp, \mu, \nu)
\,,\nn\\
\nu \frac{\df}{\df \nu}\, S_i(k^+,\vec k_\perp,\mu,\nu) &= -2 \int\! \df^2 \vec k_\perp'\, \gamma_\nu^i(\vec k_\perp - \vec k_\perp',\mu)\, S_i(k^+,\vec k_\perp',\mu,\nu)
\,,\nn\\
\gamma^i_S(\mu, \nu)
&= 4 \Gamma_\cusp^i(\al_s)\, \ln \frac{\mu}{\nu} -2\big[\gamma_H^i(\alpha_s) + \gamma_J^i(\alpha_s)\big]
\,,\end{align}
with $\ga_\nu^i$ given in \eq{ga_TMD_B}.

  The anomalous dimensions of the $n$-collinear-soft function and $\bar n$-collinear-soft function are identical. Using the $\mu$ and $\nu$ independence of the cross section in \eq{DY_fact_SCETonetwo}, they are constrained by consistency to be
  \begin{align} \label{eq:gamma_CS}
\mu\frac{\df}{\df\mu}\, \CS_i(k^+,\vec k_\perp,  \mu,\nu)
&= \int_0^{k^+}\! \df {k'}^+\, \gamma_\CS^i(k^+\! - {k'}^+, \mu,\nu)\, \CS_i({k'}^+, \vec k_\perp,\mu,\nu)
\,, \nn \\  
\nu \frac{\df}{\df \nu}\, \CS_i(k^+,\vec k_\perp,\mu,\nu) &= - \int\! \df^2 \vec k_\perp'\, \gamma_\nu^i(\vec k_\perp - \vec k_\perp',\mu)\, \CS_i(k^+,\vec k_\perp',\mu,\nu)
    \nn \\  
    \ga^i_{\CS}(k^+,\mu,\nu) &= -2 \Gamma_\cusp^i(\al_s)
    \bigg[ \frac{1}{\mu} \cL_0\Big(\frac{k^+}{\mu}\Big) + 
   \ln \frac{\nu}{\mu}\, \de(k^+)\bigg] 
  \,.\end{align}

\subsection{NLL Cross Section}
\label{sec:nll}

At NLL, the cross section is generated by evolving the tree-level functions from their natural scale\footnote{The inclusion of the factor of $-\img$ in the hard scale $\mu_H$ follows from  \eq{C_RGE} and allows us to resum a series of $\pi^2$-terms~\cite{Parisi:1979xd, Sterman:1986aj, Magnea:1990zb, Eynck:2003fn}, thereby improving convergence.} 
  \begin{align} \label{eq:canonical}
    \mu_H &= - \img Q
    \,, \nn \\    
    \mu_B &= p_T\,, \quad \nu_B = Q 
    \,, \nn \\
    \mu_\CS &= p_T\,, \quad \nu_\CS = p_T^2 / \Tau
    \,, \nn \\
    \mu_S &= \Tau 
  \,.\end{align}
to a common scale using the RG equations in \sec{renormalization}.  
Evolving all functions to the collinear-soft scale $(\mu_{\CS}, \nu_{\CS})$, using results from refs.~\cite{Balzereit:1998yf, Neubert:2004dd, Fleming:2007xt, Ligeti:2008ac,Manohar:2012jr,Chiu:2012ir,Almeida:2014uva}, we obtain
  \begin{align} \label{eq:nll_res}
    &\int_0^{p_T}\! \df p_T'\,\int_0^{\Tau}\! \df \Tau'\, \frac{\df \si}{\df Q^2\,\df Y\, \df p_T'\, \df \Tau'}
    \nn \\ & \quad
    = \sum_q \hat \si_q^\zero\,  \big[ f_q(x_1,\mu_B) f_{\bar q}(x_2,\mu_B) + f_{\bar q}(x_1,\mu_B) f_q(x_2,\mu_B) \big]
    \nn \\ & \qquad \times    
 \frac{\Gamma(1-\eta_B)\,e^{\text{Re}(K_H) + K_B + K_S - 2\gamma_E\, \eta_B - \gamma_E\, \eta_S}}{\Gamma(1+\eta_B)\Gamma(1+\eta_S)}
 \bigg|\Bigl(\frac{-Q^2-\img 0}{\mu_H^2}\Bigr)^{\eta_H}\bigg|
\Bigl( \frac{p_T}{\mu_B} \Bigr)^{2\eta_B}
\Bigl( \frac{\Tau}{\mu_S} \Bigr)^{\eta_S}
\,,\end{align}
where $x_{1,2} = (Q/\Ecm) e^{\pm Y}$ and $\text{Re}(\dots)$ denotes the real part of a complex number. 
The evolution kernels are given by
\begin{align}
K_H(\mu_H,\mu_\CS) &= -4 K_\Gamma^q(\mu_H,\mu_\CS) + 2K_{\gamma_H^q}(\mu_H,\mu_\CS)
\,, \quad &
\eta_H(\mu_B,\mu_\CS) &= 2 \eta^q_{\Gamma}(\mu_B,\mu_\CS)
\,,\nn \\
K_B(\nu_B,\nu_\CS) &= 2 \ga_\nu^q(\al_s) \ln \Big(\frac{\nu_\CS}{\nu_B} \Big)
\,, \quad &
\eta_B(\nu_B,\nu_\CS) &= -2 \eta^q_{\Gamma}(\nu_B,\nu_\CS)
\,,\nn\\
K_S(\mu_S,\mu_\CS) &= -4 K_\Gamma^q(\mu_S,\mu_\CS) + K_{\gamma_S^q}(\mu_S,\mu_\CS)
\,, \quad &
\eta_S(\mu_S,\mu_\CS) &= 4 \eta^q_{\Gamma}(\mu_S,\mu_\CS)
\,,\end{align}
in terms of functions given in \app{RGE}.
Since $\mu_B = \mu_\CS$ there is no $\mu$-evolution for the beam functions. Because the scale of $\al_s$ in the $\nu$-evolution is $\mu$, the evolution of the non-cusp $\nu$-anomalous dimension takes the simpler form shown in $K_B$.

It is worth emphasizing that \eq{nll_res} continuously merges with the \SCETa and \SCETb boundaries. This is no longer automatically achieved at NNLL, but can still be arranged, as discussed in the next section. We also stress that \eq{canonical} represents a naive choice of scales as these do no smoothly turn off at the boundaries leading to a discontinuity in the derivative of the cross section (see also the discussion around \eq{deriv_constr}). This will be remedied by using profile functions~\cite{Ligeti:2008ac,Abbate:2010xh} in ref.~\cite{numerics}, where a full analysis at NNLL will be presented.

\section{Matching the Effective Theories}
\label{sec:matching}

We now show that the continuous description of the cross section across the \SCETa, \SCETab and \SCETb regions discussed in \sec{nll} can naturally be extended to all orders. Specifically, in the \SCETab region of phase space,
\begin{align} \label{eq:match_allorders}
  \cI_{ij}(t,x,\vec k_\perp,\mu) &= \int\! \df^2 \vec k_\perp'\, \cI_{ij}(x,\vec k_\perp',\mu,\nu)\, \CS(t/p^-,\vec k_\perp - \vec k_\perp',\mu,\nu)
 \,, \nn \\
S(k^+,\vec k_\perp,\mu,\nu) &= \int\! \df^2 \vec k_\perp' \int\! \df k'^+\, \df k''^+\,
  \CS(k'^+,\vec k_\perp',\mu,\nu)   \CS(k''^+,\vec k_\perp \!- \vec k_\perp',\mu,\nu)
\nn \\ & \quad \times 
 S(k^+ \!- k'^+ \!- k''^+,\mu)
\,,\end{align}
up to power corrections of $\ord{\vks/t}$ and $\ord{(k^+)^2/\vec k_\perp^2}$, respectively.
This follows directly from the consistency of the factorization theorems in \sec{fact_formula}: When the resummation is turned off, i.e.~a common renormalization scale is chosen for all functions in the factorization theorem, the \SCETa and \SCETb factorization theorems simply produce the full \emph{fixed-order} cross section up to power corrections. As the \SCETab regime involves an additional expansion, its fixed-order cross section can be obtained from either. Due to the many common ingredients between the \SCETab, \SCETa and \SCETb factorization theorems, this then implies \eq{match_allorders}.

We now restrict our attention to NNLL, for which \eq{match_allorders} reduces to
\begin{align} \label{eq:match}
  \cI_{qq}^\one(t,x,\vec k_\perp,\mu) &=  \de(t)\, \cI_{qq}^\one(x,\vec k_\perp,\mu,\nu) + \de(1-x)\, \CS^\one(t/p^-,\vec k_\perp,\mu,\nu)
  \nn \\
  \cI_{qg}^\one(t,x,\vec k_\perp,\mu) &= \de(t)\, \cI_{qg}^\one(x,\vec k_\perp,\mu)
 \,, \nn \\
S^\one(k^+,\vec k_\perp,\mu,\nu) &=  \frac{1}{\pi}\,\de(\vks) S^\one(k^+,\mu) + 2 \CS^\one(k^+,\vec k_\perp,\mu,\nu) 
\,.\end{align}
The first equations are valid up to corrections of $\ord{\vks/t}$, whereas the last one holds exactly for $k^+ < |\vec k_\perp|$.
This naturally suggests the following procedure for patching together the cross section at NNLL,\footnote{This has a natural generalization beyond NNLL in Fourier/Laplace space, where one can take the full inverse of $\CS$ rather than the expanded version employed here.}
\begin{align} 
\frac{\df^4 \si}{\df Q^2\, \df Y\, \df p_T^{\,2}\, \df \Tau}
&=  \sum_{q}\,\hat \sigma_{q}^0 \, H(Q^2, \mu)  \!\int\!\df t_1\, \df t_2  \int\! \df^{2} \vec k_{1\perp}\, \df^{2} \vec k_{2\perp}\, \df^{2} \vec k_{1\perp}^{\rm cs}\, \df^{2} \vec k_{2\perp}^{\rm cs}\, \df^2 \vec k_\perp
\int\! \df k_1^+\, \df k_2^+\, \df k^+
\nn\\ &\quad \times
\Big[ B_q(t_1,x_1, \vec k_{1\perp}, \mu) - \CS^\one\bigl(t_1 e^{-Y}\!\!/Q,\vec k_{1\perp},\mu,\nu \bigr)\Big]
\CS\bigl(k_1^+,\vec k_{1\perp}^{\rm cs},\mu,\nu \bigr) 
\nn\\ &\quad \times
\Big[ B_{\bar q}(t_2,x_2, \vec k_{2\perp},\mu) - \CS^\one\bigl(t_2 e^Y\!\!/Q,\vec k_{2\perp},\mu,\nu \bigr)\Big]
\CS \bigl(k_2^+,\vec k_{2\perp}^{\rm cs},\mu,\nu \bigr)
\nn \\ & \quad \times
\Big[S(k^+, \vec{k}_{\perp}, \mu, \nu) - 2\CS^\one(k^+, \vec{k}_{\perp}, \mu, \nu)\Big]
\de\Big(\Tau \!-\! \frac{e^{-Y} t_1 \!+\! e^Y t_2}{Q} \!-\! k_1^+ \!-\! k_2^+ \!-\! k^+\Big)
\nn \\ & \quad \times
\de\big(p_T^{\,2} - \abs{\vec k_{1\perp}+ \vec k_{2\perp}+\vec k_{1\perp}^{\rm cs}+ \vec k_{2\perp}^{\rm cs} + \vec k_\perp}^{2} \big)  + (q\leftrightarrow \bar q)
\,.\end{align}
Here the $\CS^{\one}$-term subtracted from the beam functions (soft function) are evaluated at the beam (soft) scale. From \eq{match} it follows that this reproduces the \SCETa, \SCETab and \SCETb factorization theorems in eqs.~\eqref{eq:DY_fact_SCETone}, \eqref{eq:DY_fact_SCETonetwo} and \eqref{eq:DY_fact_SCETtwo}, up to power corrections.

We now derive \eq{match}, using cumulants to avoid subtleties related to distributions.
Starting with the boundary between \SCETa and \SCETab, 
\begin{align}
  \int_0^t\! \df t' \int_0^{\vks} \! \df \vkps\, \cI_{qg}^\one(t',x,\vec k_\perp',\mu)
  &=
  \frac{\al_s T_F}{2\pi^2}\,\bigg[\ln \min \Big\{\frac{(1-x)t}{x \mu^2}, \frac{\vec k_\perp^{\,2}}{\mu^2}\Big\}\,
  P_{qg}(x) + 2x(1-x)  \bigg] 
\,,\nn \\
  \int_0^{\vks} \! \df \vkps\, \cI_{qg}^\one(x,\vec k_\perp',\mu,\nu)
  &=
\frac{\al_s T_F}{2\pi^2} \bigg[ \ln \Big(\frac{\vec k_\perp^{\,2}}{\mu^2}\Big) P_{qg}(x) 
+ 2 x(1-x) \bigg]
\,.\end{align}
We thus obtain the second line in \eq{match} for $0 < x < 1- \de$, where 
\begin{align} \label{eq:delta}
\de = \frac{\vks}{t + \vks}
\,.\end{align}
In the \SCETab region of phase space, the size $\de$ of the remaining interval $1-\de \leq x \leq 1$ is parametrically small, implying that the contribution from this region to the cross section is power suppressed. 

Similarly, we find that for $0 < x < 1- \de$ the first line of \eq{match} is satisfied,
\begin{align} \label{eq:Iqq_check}
  \int_0^t\! \df t' \int_0^{\vks} \! \df \vkps\, \cI_{qq}^\one(t',x,\vec k_\perp',\mu)
  &=
  \frac{\al_s C_F}{2\pi^2}\,\bigg[\ln \Big(\frac{\vec k_\perp^{\,2}}{\mu^2}\Big)\,
  P_{qq}(x) + 1-x  \bigg] 
  =   \int_0^{\vks} \! \df \vkps\, \cI_{qq}^\one(x,\vec k_\perp',\mu,\nu)
\,.\end{align}
Although $1-\de \leq x \leq 1$ is again parametrically small, the integral over this region is not, due to the presence of delta functions and plus distributions at $x = 1$,
\begin{align}
  \int_0^t\! \df t' \int_0^{\vks} \! \df \vkps\, \int_{1-\de}^1\! \df x\, \cI_{qq}^\one(t',x,\vec k_\perp',\mu)
  &=
  \frac{\al_s C_F}{2\pi^2}\,\bigg[\ln^2 \Big(\frac{\de\, t}{\mu^2}\Big) - \frac{\pi^2}{6} + \ord{\de} \bigg]
\,,\nn \\
  \int_0^{\vks} \! \df \vkps\, \int_{1-\de}^1\! \df x\, \cI_{qq}^\one(x,\vec k_\perp',\mu)
  &=
\frac{\al_s C_F}{2\pi^2} \bigg[ 2 \ln \Big(\frac{\vec k_\perp^{\,2}}{\mu^2}\Big)\,\ln \Big(\frac{\de\, p^-}{\nu}\Big) + \ord{\de} \bigg]
\,.\end{align}
The mismatch is captured by the collinear-soft function
\begin{align}
  \int_0^{t/p^-}\!\!\!\! \df k^+ \!\! \int_0^{\vks} \! \df \vkps\, \CS^\one(k^+,\vec k_\perp',\mu,\nu)
  &= \frac{\al_s C_F}{2\pi^2} \bigg[\!-\! \ln^2 \Big(\frac{\vks}{\mu^2}\Big) 
  \!+\! 2 \ln \Big(\frac{\vks}{\mu^2}\Big) \Big(
  \ln \Big(\frac{t}{\mu p^-}\!\Big) \!+\! \ln \frac{\nu}{\mu}\Big) \!-\! \frac{\pi^2}{6}\bigg]
\,,\end{align}
up to a power suppressed contribution
\begin{align}
  & \int_0^{\vks} \! \df \vkps\! \int_{1-\de}^1\! \df x\, \bigg[ \int_0^t\! \df t'\, \cI_{qq}^\one(t',x,\vec k_\perp',\mu) 
- \cI_{qq}^\one(x,\vec k_\perp',\mu) \bigg]
  -  \int_0^{t/p^-}\!\!\!\! \df k^+ \!\! \int_0^{\vks} \! \df \vkps\, \CS^\one(k^+,\vec k_\perp',\mu,\nu) 
 \nn \\ & \quad 
 =  \frac{\al_s C_F}{2\pi^2} \ln^2 \frac{\de t}{\vks} + \ord{\de} = \ord{\de}
\,.\end{align}
Note that in the last line it important that $\de$ is not arbitrary but given by \eq{delta}. Combined with \eq{Iqq_check}, this establishes the first line of \eq{match}.

Lastly, we consider the boundary between \SCETab and \SCETb, which involves the following ingredients
\begin{align} \label{eq:scetb_matching}
  \int_0^{k^+}\!\!\! \df k'^+ \!\! \int_0^{\vks} \!\! \df \vkps\, S^\one(k'^+,\vec k_\perp',\mu,\nu)
  &= \frac{\al_s C_F}{2\pi^2} \bigg[ 
  - 2 \ln^2 \Big(\frac{\vks}{\mu^2}\Big)
  + 4 \ln \Big(\frac{\vks}{\mu^2}\Big) \ln \Big(\frac{k^+}{\mu}\Big)
   -4 \ln^2 \Big(\frac{k^+}{\mu}\Big)
  \nn \\ & \quad
   + 4 \ln \Big(\frac{\vks}{\mu^2}\Big) \ln \frac{\nu}{\mu} - \frac{\pi^2}{6}
   + \theta((k^+)^2 - \vks) \ln^2 \Big(\frac{\vks}{(k^+)^2} \Big)
  \bigg]
\,, \nn \\
  \int_0^{k^+}\!\! \df k'^+\, S^\one(k'^+,\mu)
  &= \frac{\al_s C_F}{2\pi} \bigg[ -4 \ln^2 \Big(\frac{k^+}{\mu}\Big) + \frac{\pi^2}{6}\bigg]
\,, \nn \\
  \int_0^{k^+}\!\!\! \df k'^+\!\! \int_0^{\vks} \!\! \df \vkps\, \CS^\one(k'^+,\vec k_\perp',\mu,\nu)
  &= \frac{\al_s C_F}{2\pi^2} \bigg[\!-\! \ln^2 \Big(\frac{\vks}{\mu^2}\Big) 
  \!+\! 2 \ln \Big(\frac{\vks}{\mu^2}\Big) \Big(
  \ln \Big(\frac{k^+}{\mu}\Big) \!+\! \ln \frac{\nu}{\mu}\Big) \!-\! \frac{\pi^2}{6}\bigg]
\,.\end{align}
It is straightforward to verify that for $k^+ < k_\perp$ this satisfies the last line in \eq{match}.

\section{NLO Cross Section}
\label{sec:NLO}

In this section we determine the NLO cross section for $Z+0$ jet production, differential in
the invariant mass $Q^2$, the rapidity $Y$ and $p_T$ of the $Z$ and beam thrust $\Tau$. We start by collecting the relevant ingredients in \sec{NLO_ingr}, check the cancellation of IR divergences in \sec{IR_div} and present the final result in \sec{NLO_result}. In \sec{NLO_resum} we verify that this agrees with \SCETa, \SCETab and \SCETb, up to power corrections. This provides an important cross check of our formalism. We will match our resummed prediction onto these fixed-order corrections in ref.~\cite{numerics}.

\subsection{Ingredients}
\label{sec:NLO_ingr}

   The partonic cross section for the one-loop real and virtual corrections in $\overline{\text{MS}}$ are given by
   \begin{align} \label{eq:matrix_elem}
    \hat \si^\one_{q,R} &= Q^2\,\hat \si_q^\zero\, 8\pi\, \al_s C_F\,\Big(\frac{e^{\ga_E}\mu^2}{4\pi}\Big)^\eps\, \frac{1}{s_{q\bar qg}}\Big[ \frac{s_{qg}}{s_{\bar qg}} + \frac{s_{\bar qg}}{s_{qg}} + 2 \frac{s_{q\bar q}s_{q\bar qg}}{s_{qg}s_{\bar qg}} - \eps \Big (2 + \frac{s_{qg}}{s_{\bar qg}} + \frac{s_{\bar qg}}{s_{qg}} \Big)  \Big]
    \,, \nn \\
    \hat \si^\one_{q,V} &= Q^2\,\hat \si_q^\zero\, \frac{\al_s C_F}{\pi}\, \Big(\frac{\mu^2}{s_{q\bar qg}}\Big)^\eps \Big[-\frac{1}{\eps^2} - \frac{3}{2\eps} -4 + \frac{7\pi^2}{12} + \ord{\eps} \Big]
   \,.\end{align}
   The Lorentz invariants that enter here are defined as
   \begin{align}
   s_{ij} = (p_i + p_j)^2 = 2 p_i \sdt p_j
   \,, \qquad
   s_{ijk} = (p_i + p_j + p_k)^2 = s_{ij} + s_{ik} + s_{jk}
   \,,\end{align}
   using an incoming momentum convention for $p_i$. Due to the flavor dependence of the tree-level partonic cross section $\hat \si^\zero_q$, we will for simplicity restrict ourselves to a single quark flavor. The full cross section can be obtained by summing over quark flavors.

We now discuss kinematics and phase space. The incoming partons have momenta 
\begin{align}
   p_1 = (x_1 \Ecm, 0,0)
\,, \qquad   
p_2 = (0,x_2 \Ecm,0)
\,,\end{align}
in $(-,+,\perp)$ light-cone coordinates (see \eq{lc}), with $x_{1,2}$ the momentum fractions of the partons with respect to the colliding hadrons. At LO the final state consists of a $Z$ boson with momentum $q^\mu$, and the phase space integral yields
  \begin{align} \label{eq:LO_PS}
  \int \df\Phi_{ij}^\zero &=\int\! \frac{\df x_1}{x_1}\, \frac{\df x_2}{x_2}\, f_i(x_1,\mu) f_j(x_2,\mu) \int\! \frac{\df^d q}{(2\pi)^d}\, (2\pi)^d \de(q - p_1 - p_2)
  \nn \\ & \quad \times
  \de(Q^2 - q^2)\, \de\Big[Y - \tfrac12 \ln \Big(\frac{q^-}{q^+}\Big) \Big]\, \de(p_T^2 - \vec q_\perp^{\,2})\, \de(\Tau)\, 
    \nn \\ & 
    = \frac{1}{Q^2}\, \de(p_T^2)\, \de(\Tau) \, f_i\Big( \frac{Q}{\Ecm}\, e^Y, \mu \Big)\, f_j\Big(\frac{Q}{\Ecm}\, e^{-Y}, \mu \Big)
  \,.\end{align}
  At this order, the momentum fractions $x_{1,2}$ and the momentum of the $Z$ are thus
\begin{align}
x_1 = \frac{Q}{\Ecm}\, e^Y
\,, \qquad
x_2 = \frac{Q}{\Ecm}\, e^{-Y}
\,, \qquad
q = (Q e^Y, Q e^{-Y},0)
\,.\end{align}

  At NLO, there is an additional massless parton that the $Z$-boson can recoil against. To be consistent with \eq{matrix_elem}, we use an incoming  convention for the momentum $p_3$ of this parton.
  Assuming for simplicity that this parton goes into the right hemisphere, $-p_3^+ < -p_3^-$, the phase space is given by
  \begin{align} \label{eq:NLO_PS}
    \int \df \Phi^\one_{ij,\text{R}} &= \int\! \frac{\df x_1}{x_1}\, \frac{\df x_2}{x_2}\, f_i(x_1,\mu) f_j(x_2,\mu) \! \int\! \frac{\df^d q}{(2\pi)^d} \! 
    \int\! \frac{\df^d p_3}{(2\pi)^d}\, \theta(-p_3^0)\, 2\pi\de(p_3^2)\, (2\pi)^d \de(q \!-\! p_1 \!-\! p_2 \!-\! p_3)  
     \nn \\ & \quad \times
    \de(Q^2 - q^2)\, 
    \de\Big[Y - \tfrac12 \ln \Big(\frac{q^-}{q^+}\Big) \Big]\, \de(p_T^2 - \vec q_\perp^{\,2})\,
     \theta(p_3^+ - p_3^-)\, \de(\Tau + p_3^+)\, 
    \nn \\ & 
    =  \frac{1}{(4\pi)^{2-\eps} \Ga(1-\eps)}\, 
    \frac{\theta(p_T - \Tau)}{p_T^{2\eps} \Big(\sqrt{Q^2+p_T^2}\, e^Y\, \Tau + p_T^2\Big)\Big(\sqrt{Q^2 +p_T^2}\, e^{-Y} +\Tau\Big) } 
   \nn \\ & \quad \times       
   f_i\bigg( \frac{\sqrt{Q^2+p_T^2}\, e^Y}{\Ecm} + \frac{p_T^2}{\Ecm \Tau},\mu \bigg)
    f_j\bigg(\frac{\sqrt{Q^2 +p_T^2}\, e^{-Y}}{\Ecm}  + \frac{\Tau}{\Ecm},\mu \bigg)
  \,.\end{align}
    The contribution for the other hemisphere $\df \Phi^\one_{ij,\text{L}}$ can be obtained in a similar manner.
  From this we can read off
  \begin{align} \label{eq:nlo_kin}
    x_1 &= \frac{\sqrt{Q^2+p_T^2}\, e^Y}{\Ecm} + \frac{p_T^2}{\Ecm \Tau}
    \,, &
    x_2 &= \frac{\sqrt{Q^2 +p_T^2}\, e^{-Y}}{\Ecm}  + \frac{\Tau}{\Ecm}
    \,, \nn \\ 
    q &= \Big(\sqrt{Q^2 + p_T^2}\, e^Y, \sqrt{Q^2 + p_T^2}\, e^{-Y}, p_T\Big)
    \,, &
    p_3 &= \Big(\frac{-p_T^2}{\Tau},-\Tau,p_T\Big)
  \,.\end{align}
  The (irrelevant) azimuthal angle in the transverse plane is not fixed by the measurement.
  It is straightforward to evaluate the invariants in \eq{matrix_elem} in terms of \eq{nlo_kin}. For $q \bar q \to Z g$,
  \begin{align}
    s_{q \bar q} =  x_1 x_2 \Ecm^2
    \,, \quad
    s_{qg} = - x_1 \Ecm \Tau
    \,, \quad
    s_{\bar q g} = - x_2 \Ecm \, \frac{p_T^2}{\Tau}
  \,.\end{align}
The other cases can be obtained by permutations. For $g q \to Z q$ we have
  \begin{align}
    s_{q\bar q} = - x_2 \Ecm \, \frac{p_T^2}{\Tau}
    \,, \quad
    s_{qg} =  x_1 x_2 \Ecm^2
    \,, \quad
    s_{\bar q g} = - x_1 \Ecm \Tau
  \,,\end{align}
  and for $q g \to Z q$ we have
  \begin{align}
    s_{q\bar q} = - x_1 \Ecm \Tau
    \,, \quad
    s_{qg} =  x_1 x_2 \Ecm^2
    \,, \quad
    s_{\bar q g} = - x_2 \Ecm \, \frac{p_T^2}{\Tau}
  \,.\end{align}

   Lastly, there is the NLO contribution from the PDFs, which consists of pure IR poles in dimensional regularization. This can be {\emph {effectively}} described as
    \begin{align}
       f_q^\one(x,\mu) = \frac{\al_s}{2\pi}\, \frac{1}{\eps}\, \sum_j \int_x^1\!\frac{\df x'}{x'}\, C_j P_{qj}\Big(\frac{x}{x'}\Big) f_j(x',\mu)
  \,,\end{align}
   where the color factor $C_j$ is $C_F$ ($T_F$) for $j=q$ ($j=g$) and the splitting functions are
\begin{align} \label{eq:P_def}
P_{qq}(z)
= (1+z^2) \cL_0(1-z) + \frac{3}{2}\,\delta(1-z)
\,, \qquad
P_{qg}(z) = (1-z)^2+ z^2
 \,.\end{align}

\subsection{Cancellation of IR divergences}
\label{sec:IR_div}
  
  In this section we combine these ingredients and verify the cancellation of IR divergences.
  We assume $p_T , \Tau \ll Q$ to simplify the calculation, though we do not restrict to any particular \emph{relative} hierarchy between $p_T$ and $\Tau$. This leads to 
   \begin{align} 
    \hat \si_{q\bar q \to Zg} &= Q^2 \hat \si_q^\zero\, 8\pi\, \al_s C_F\,\Big(\frac{e^{\ga_E}\mu^2}{4\pi}\Big)^\eps\, \Big[ \frac{2}{p_T^2} + 
    (1-\eps)\, \frac{p_T^2}{x_1 \Ecm(x_1 \Ecm \Tau - p_T^2) \Tau}  \Big] \Big[1 + \ORd{\frac{\Tau}{Q},\frac{p_T^2}{Q^2}}\Big]
    \nn \\ 
    \hat \si_{gq \to Zq} &= Q^2 \hat \si_q^\zero\, 8\pi\, \al_s T_F\,\Big(\frac{e^{\ga_E}\mu^2}{4\pi}\Big)^\eps\, \Big[ \frac{1}{x_1 \Ecm \Tau - p_T^2} - \frac{2}{1-\eps}\, \frac{p_T^2}{x_1^2 \Ecm^2 \Tau^2}  \Big]  \Big[1 + \ORd{\frac{\Tau}{Q},\frac{p_T^2}{Q^2}}\Big]
    \nn \\ 
    \hat \si_{qg \to Zq} &= Q^2 \hat \si_q^\zero\, 8\pi\, \al_s T_F\,\Big(\frac{e^{\ga_E}\mu^2}{4\pi}\Big)^\eps\, \frac{x_1 \Tau^2}{x_2 p_T^2(x_1 \Ecm \Tau - p_T^2)}   \Big[1 + \ORd{\frac{\Tau}{Q},\frac{p_T^2}{Q^2}}\Big]
  \,. \end{align}
   For $qg \to Zq$ and $gq \to Zq$ there is a fermion minus sign from crossing \eq{matrix_elem} and we have taken into account that we need to  average over incoming gluon polarizations and colors instead of quark spins and colors, resulting in the overall factor $2 N_c/[(2-2\eps)(N_c^2-1)]$.
The phase space in \eq{NLO_PS} simplifies as well
\begin{align} 
 \int \df \Phi_{ij,R}^\one &=
    \frac{1}{(4\pi)^{2-\eps} \Ga(1-\eps)}\, 
    \frac{\theta(p_T - \Tau)}{Q( Q \Tau + p_T^2 e^{-Y}) p_T^{2\eps}} \,
    f_i\bigg( \frac{Q\, e^Y}{\Ecm} + \frac{p_T^2}{\Ecm \Tau},\mu \bigg)\, 
    f_j\bigg(\frac{Q\, e^{-Y}}{\Ecm},\mu \bigg)
\,.\end{align}
  
   To avoid subtleties related to distributions, we  calculate the cumulative cross section in $p_T$ and $\Tau$,
    \begin{align} \label{eq:si_ir}
    &\int_0^{p_T^2}\! \df p_T'^{\,2} \int_0^{\Tau}\! \df \Tau'\, \frac{\df^4 \si_q^\one}{\df Q^2\, \df Y\, \df p_T'^{\,2}\, \df \Tau'}
    \nn \\ & \quad
    = \int_0^{p_T^2}\! \df p_T'^{\,2} \int_0^{\Tau}\! \df \Tau'\, \bigg\{\int \df \Phi_{q \bar q}^\zero\, \bigg[\frac12\,\hat \si_{q,V}^\one + \hat \si_q^\zero\, \frac{f_q^\one(x_1,\mu)}{f_q(x_1,\mu)} \bigg]  + \int \df \Phi_{q \bar q,R}^\one\, \hat \si_{q \bar q \to Zg} 
    \nn \\ & \qquad    
    + \int \df \Phi_{q g,R}^\one\, \hat \si_{q g \to Zq} + \int \df \Phi_{g q,R}^\one\, \hat \si_{g q \to Zq} 
    + (x_1 \lra x_2) + (q \lra \bar q) \bigg\} 
     \nn \\ & \quad    
    = \hat \si_q^\zero\, f_q(x_1,\mu) f_{\bar q}(x_2,\mu)\, \frac{\al_s}{2\pi} \bigg(C_F \, \Big(\frac{\mu^2}{Q^2}\Big)^\eps \Big[- \frac{1}{\eps^2} - \frac{3}{2\eps} - 4 + \frac{7\pi^2}{12} + \ord{\eps} \Big]
    \nn \\ & \qquad
    + \frac{1}{\eps} \sum_j \int_{x_1}^1 \frac{\df x_1'}{x_1'}\, \frac{f_j(x_1',\mu)}{f_q(x_1,\mu)}\, C_j P_{qj}\Big(\frac{x_1}{x_1'}\Big)
    \nn \\ & \qquad
    - \frac{1}{\eps} \int_{x_1}^1\! \frac{\df x_1'}{x_1'}\, \bigg\{\frac{f_q(x_1',\mu)}{f_q(x_1,\mu)}\, C_F\Big[ \frac{2x_1}{(x_1' - x_1)^{1+2\eps}} + (1-\eps) \,\frac{(x_1' - x_1)^{1-2\eps}}{x_1'}\Big]
    \nn \\ & \qquad
     + \frac{f_g(x_1',\mu)}{f_q(x_1,\mu)}\, T_F\, \Big[\frac{1}{(x_1'-x_1)^{2\eps}} - \frac{2}{1-\eps}  \frac{x_1(x_1'-x_1)^{1-2\eps}}{x_1'^2} \Big]\bigg\}  \frac{e^{\eps \ga_E}}{\Ga(1-\eps)}\, \Big(\frac{\mu^2}{\Ecm^2}\Big)^\eps
    \nn \\ & \qquad \times 
     \bigg[\min \bigg\{1, \frac{\Tau}{(x_1' - x_1) \Ecm}, \frac{p_T^2}{(x_1'-x_1)^2\Ecm^2} \bigg\}\bigg]^{-\eps}
    \nn \\ & \qquad
    + \int_{x_1}^1\! \frac{\df x_1'}{x_1'}\, \frac{f_q(x_1',\mu) f_g(x_2,\mu)}{f_q(x_1,\mu) f_{\bar q}(x_2,\mu)}\, T_F\, \frac{x_1'}{x_2 (x_1'-x_1)^{2\eps}}  \frac{e^{\eps \ga_E}}{\Ga(2-\eps)}\, \Big(\frac{\mu^2}{\Ecm^2}\Big)^\eps
    \nn \\ & \qquad \times 
     \bigg[\min \bigg\{1, \frac{\Tau}{(x_1' - x_1) \Ecm}, \frac{p_T^2}{(x_1'-x_1)^2\Ecm^2} \bigg\}\bigg]^{1-\eps}     
   \bigg)  + (x_1 \lra x_2) + (q \lra \bar q) 
    \,, \end{align}
    where in the final expression we used the shorthand notation 
\begin{align}
x_1 = \frac{Q}{\Ecm}\, e^Y
\,, \qquad
x_2 = \frac{Q}{\Ecm}\, e^{-Y}
\,,\end{align}
which should not to be confused with \eq{nlo_kin}. The contribution from $\df \Phi_L$ is included through $(x_1 \lra x_2)$.

We obtained \eq{si_ir} by first rewriting the $p_T'^{\,2}$ integral in terms of $x'$
\begin{align}
p_T'^{\,2}= (x_1' - x_1) \Ecm\, \Tau'
\,.\end{align}
such that
\begin{align}
   \int_0^{p_T^2}\! \df p_T'^{\,2}
   = \int_{x_1}^1 \! \df x_1'\, \Ecm\, \Tau'\, \theta\big(p_T^2 - (x_1'-x_1) \Ecm \Tau'\big)
\,.\end{align}
For the subsequent $\Tau'$ integral we find
\begin{align}
&\int_0^{\Tau}\! \df \Tau'\, \frac{\theta\big(p_T^2 - (x_1'-x_1) \Ecm \Tau'\big) \theta\big((x_1'-x_1)\Ecm - \Tau')}{(\Tau')^{1+\eps}}
\nn \\ & \quad 
= - \frac{1}{\eps}\, \frac{\theta(x_1'-x_1)}{(x_1' - x_1)^\eps \Ecm^\eps}\, \bigg[\min \bigg\{1, \frac{\Tau}{(x_1' - x_1) \Ecm}, \frac{p_T^2}{(x_1'-x_1)^2\Ecm^2} \bigg\}\bigg]^{-\eps}
\,,\end{align}
and similarly for the term that goes like $(\Tau')^{-\eps}$.
The cancellation of IR divergences becomes clear once we use the following expressions to expand in $\eps$,
\begin{align}
\frac{2x_1}{(x_1' \!-\! x_1)^{1+2\eps}} + (1\!-\!\eps) \,\frac{(x_1' \!-\! x_1)^{1-2\eps}}{x_1'} &= \Big(\!-\! \frac{1}{\eps} \!-\! \frac{3}{2} \!+\! 2 \ln x_1\Big) \de\Big(1 \!-\! \frac{x_1}{x_1'}\Big) + P_{qq}\Big(\frac{x_1}{x_1'}\Big)  + \ord{\eps}
\,, \nn \\ 
\frac{1}{(x_1'-x_1)^{2\eps}} - \frac{2}{1-\eps}  \frac{x_1(x_1'-x_1)^{1-2\eps}}{x_1'^2} &= P_{qg}\Big(\frac{x_1}{x_1'}\Big)  + \ord{\eps}
\,,\end{align}
which follow from \eq{plusexp}. Note that the $\ln x_1$ term on the first line and the corresponding term from $(x_1 \lra x_2)$ combine with $\ln (E^2/\mu^2)$ to give $\ln (Q^2/\mu^2) = \ln x_1 + \ln x_2 + \ln (E^2/\mu^2)$.

\subsection{Result}
\label{sec:NLO_result}
  
  We now present the cross section for $pp \to Z+0$ jets, differential in the invariant mass and rapidity of the $Z$, and with cuts on the transverse momentum of the $Z$ and on beam thrust.
  This is given by the finite $\ord{\eps^0}$ terms in \eq{si_ir}, which we rearrange into the following form
    \begin{align} \label{eq:si_fin}
    &\int_0^{p_T^2}\! \df p_T'^{\,2} \int_0^{\Tau}\! \df \Tau'\, \frac{\df^4 \si_q^\one}{\df Q^2\, \df Y\, \df p_T'^{\,2}\, \df \Tau'}
    \nn \\ & \quad
  = \hat \si_q^\zero\, f_q(x_1,\mu) f_{\bar q}(x_2,\mu)\, \frac{\al_s}{2\pi} \bigg(
    2 \bigg[ \ln^2 \Big(\frac{x_1 \Ecm}{\mu}\Big) - \ln^2 \Big(\frac{Q}{\mu}\Big) \bigg] +
    \sum_j  \int_{x_1}^1 \frac{\df z_1}{z_1}\,\frac{f_j(x_1/z_1,\mu)}{f_q(x_1,\mu)}\, C_j P_{qj}(z_1)
    \nn \\ & \qquad \times
   \ln \min \bigg\{\frac{x_1^2 \Ecm^2}{z_1^2 \mu^2}, \frac{\Tau\, x_1 \Ecm}{z_1(1\!-\!z_1) \mu^2}, \frac{p_T^2}{(1\!-\!z_1)^2\mu^2} \bigg\}
   + \int_{x_1}^1\! \frac{\df z_1}{z_1}\, \bigg\{\frac{f_q(x_1/z_1,\mu)}{f_q(x_1,\mu)}\, C_F\Big[ 2 (1\!+\!z_1^2) \cL_1(1\!-\!z_1) 
   \nn \\ & \qquad
   + \Big(- 4 + \frac{\pi^2}{2}\Big) \de(1-z_1)
   + 1-z_1 \Big]
     + \frac{f_g(x_1/z_1,\mu)}{f_q(x_1,\mu)}\, T_F\, \Big[2 P_{qg}(z_1) \ln (1-z_1)+2z_1(1-z_1) \Big]\bigg\} 
    \nn \\ & \qquad
    + \int_{x_1}^1\! \frac{\df z_1}{z_1}\, \frac{f_q(x_1/z_1,\mu) f_g(x_2,\mu)}{f_q(x_1,\mu) f_{\bar q}(x_2,\mu)}\, T_F\, \frac{x_1}{z_1 x_2}\,
   \min \bigg\{1, \frac{z_1 \Tau}{x_1(1-z_1) \Ecm}, \frac{z_1^2 p_T^2}{x_1^2 (1 - z_1)^2\Ecm^2} \bigg\}
   \bigg)  
   \nn \\ & \qquad
   + (x_1 \lra x_2) + (q \lra \bar q)
    \,.\end{align}
Here we changed variables to $z_1 = x_1/x_1'$.

\subsection{Comparison to Resummed Predictions}
\label{sec:NLO_resum}

We will now expand \eq{si_fin} in the \SCETa, \SCETab and \SCETb regions of phase space, and verify that this agrees with the predictions from factorization theorems, up to power corrections. The second-to-last line of \eq{si_fin} could never be produced by the factorization theorems, but is  power-suppressed and does not need to be considered. Since the cross section in \eq{si_fin} is a cumulative distribution, we benefit from the cumulative expressions for the ingredients of the factorization formulae in \sec{matching}.
  
The minimum in \eq{si_fin} cuts the $z_1$ interval into three regions
\begin{align} \label{eq:min_regions}
  \min \bigg\{\Big(\frac{p_1^-}{z_1 \mu}\Big)^2, \frac{\Tau\, p_1^-}{z_1(1-z_1) \mu^2}, \frac{p_T^2}{(1-z_1)^2\mu^2} \bigg\}
  = \begin{cases}
       [p_1^-/(z_1 \mu)]^2 & 1\geq z_1\geq z_a \\
       \Tau\, p_1^-/[z_1(1-z_1) \mu^2] & z_a \geq z_1 \geq z_b  \\
       p_T^2/[(1-z_1)^2\mu^2] & z_b \geq z_1 \geq 0
  \end{cases}
\end{align}
with $p_1^- = x_1 \Ecm = Q e^Y$ and boundaries at
\begin{align}
  z_a = \frac{1}{1+ \Tau/p_1^-}
  \,, \qquad
  z_b = \frac{1}{1+p_T^2/(\Tau p_1^-)}
\,.\end{align}
Because the size of the interval $1 \geq z_1 \geq z_a$ is parametrically small, $\ord{\Tau/Q}$, we only need to keep the logarithmically enhanced contributions. From the $z_1 \to 1$ behavior of the splitting functions $P_{qj}(z_1)$ in \eq{P_def}, it is clear that only the contribution from the diagonal $j=q$ term is not suppressed:
\begin{align} \label{eq:sceta_1}
    & \sum_j \int_{z_a}^1\! \frac{\df z_1}{z_1}\,\frac{f_j(x_1/z_1,\mu)}{f_q(x_1,\mu)}\,  C_j P_{qj}(z_1)\,
    \ln   \min \bigg\{\Big(\frac{p_1^-}{z_1 \mu}\Big)^2, \frac{\Tau\, p_1^-}{z_1(1-z_1) \mu^2}, \frac{p_T^2}{(1-z_1)^2\mu^2} \bigg\}
    \nn \\ & \quad
    = 2C_F \ln \Big(\frac{p_1^-}{\mu}\Big) \int_{z_a}^1\! \df z_1\,  P_{qq}(z_1) \Big[1+ \ORd{\frac{\Tau}{Q}}\Big]
    \nn \\ & \quad
    = 2C_F \ln \Big(\frac{p_1^-}{\mu}\Big) \bigg[- \ln \Big(\frac{(p_1^-)^2}{\Tau^2}\Big) + \frac{3}{2} \bigg] 
    \Big[1+ \ORd{\frac{\Tau}{Q}}\Big]
    \nn \\ & \quad
    = C_F \bigg[- 4\ln^2 \Big(\frac{p_1^-}{\mu}\Big) + 3 \ln \Big(\frac{p_1^-}{\mu}\Big) 
       + 4 \ln \Big(\frac{p_1^-}{\mu}\Big) \ln \Big(\frac{\Tau}{\mu}\Big)\bigg] \Big[1+ \ORd{\frac{\Tau}{Q}}\Big]
\end{align}

In the \SCETa region of phase space, the interval $z_a \geq z_1 \geq z_b$ is not parametrically small. We therefore do not give the boundary $z_b$ any special treatment. It is convenient to rewrite the remaining integral over $1 \geq z_1 \geq x_1$ and subtract the contribution from $1 \geq z_1 \geq z_a$. This requires us to extend $P_{qq}(z) \ln(1-z)$ to $z\to 1$, which we do as follows:
\begin{align}
  P_{qq}(z) \ln(1-z) \to (1+z^2) \cL_1(1-z)
\,.\end{align}
We thus obtain
\begin{align} \label{eq:sceta_2}
    & \sum_j \int_{x_1}^{z_a}\! \frac{\df z_1}{z_1}\,\frac{f_j(x_1/z_1,\mu)}{f_q(x_1,\mu)}\,  C_j P_{qj}(z_1)\,
    \ln \min \bigg\{\Big(\frac{p_1^-}{z_1 \mu}\Big)^2, \frac{\Tau\, p_1^-}{z_1(1-z_1) \mu^2}, \frac{p_T^2}{(1-z_1)^2\mu^2} \bigg\}
     \\ & \quad
    =  \sum_j \int_{x_1}^{1}\! \frac{\df z_1}{z_1}\,\frac{f_j(x_1/z_1,\mu)}{f_q(x_1,\mu)}\, C_j P_{qj}(z_1)\,
    \bigg[\ln \min \bigg\{\frac{\Tau\, p_1^-}{z_1\mu^2}, \frac{p_T^2}{(1-z_1)\mu^2} \bigg\} - \ln(1-z_1)\bigg]
    \nn \\ & \qquad
    - \int_{z_a}^{1}\! \df z_1\, C_F P_{qq}(z_1)\,
    \bigg[\ln \Big(\frac{\Tau\, p_1^-}{\mu^2}\Big) - \ln(1-z_1)\bigg] \Big[1+ \ORd{\frac{\Tau}{Q}}\Big]
    \nn \\ & \quad
    =  \sum_j \int_{x_1}^{1}\! \frac{\df z_1}{z_1}\,\frac{f_j(x_1/z_1,\mu)}{f_q(x_1,\mu)}\, C_j P_{qj}(z_1)\,
    \bigg[\ln \min \bigg\{\frac{\Tau\, p_1^-}{z_1\mu^2}, \frac{p_T^2}{(1-z)\mu^2} \bigg\} - \ln(1-z_1)\bigg]
    \nn \\ & \qquad
   + C_F \bigg[ 3 \ln^2 \Big(\frac{p_1^-}{\mu}\Big) - \frac{3}{2} \ln \Big(\frac{p_1^-}{\mu}\Big)  
   - 2 \ln \Big(\frac{p_1^-}{\mu}\Big) \ln \Big(\frac{\Tau}{\mu}\Big) - \ln^2 \Big(\frac{\Tau}{\mu}\Big) - \frac{3}{2} \ln \Big(\frac{\Tau}{\mu}\Big) 
   \bigg] \Big[1+ \ORd{\frac{\Tau}{Q}}\Big]
\,.\nn\end{align}
Combining eqs.~\eqref{eq:si_fin}, \eqref{eq:sceta_1} and \eqref{eq:sceta_2}, it is straightforward to verify that this agrees with the \SCETa factorization formula in \eq{DY_fact_SCETone}, using the results in \sec{matching}.

In the \SCETab and \SCETb region of phase space, the interval $z_a \geq z_1 \geq z_b$ is also parametrically small, $\ord{p_T^2/(\Tau Q)}$. In fact, for \SCETb both $z_b < z_a$ and $z_b>z_a$ are allowed. We start by assuming $z_b < z_a$,
\begin{align}
    &\sum_j \int_{z_b}^{z_a}\! \frac{\df z_1}{z_1}\,\frac{f_j(x_1/z_1,\mu)}{f_q(x_1,\mu)}\,  C_j P_{qj}(z_1)\,
    \ln \min \bigg\{\Big(\frac{p_1^-}{z_1 \mu}\Big)^2, \frac{\Tau\, p_1^-}{z_1(1-z_1) \mu^2}, \frac{p_T^2}{(1-z_1)^2\mu^2} \bigg\}
     \nn \\ & \quad
    = C_F \int_{z_b}^{z_a}\! \df z_1\,  P_{qq}(z_1)  \bigg[\ln \Big(\frac{\Tau p_1^-}{\mu^2}\Big) - \ln(1-z) \bigg]
    \Big[1+ \ORd{\frac{p_T^2}{\Tau Q}}\Big]
    \nn \\ & \quad
    = C_F \bigg\{
      - 8\ln \Big(\frac{p_1^-}{\mu}\Big) \ln \Big(\frac{\Tau}{\mu}\Big)
      - 4\ln^2 \Big(\frac{\Tau}{\mu}\Big)
      + 4 \ln \Big(\frac{p_T^2}{\mu^2}\Big) \bigg[\ln \Big(\frac{p_1^-}{\mu}\Big) + \ln \Big(\frac{\Tau}{\mu}\Big)\bigg]
      - \ln^2 \Big(\frac{p_T^2}{\mu^2}\Big)
       \bigg\} 
       \nn \\ & \qquad \times
       \Big[1+ \ORd{\frac{p_T^2}{\Tau Q}}\Big]
\,.\end{align}
The remainder is
\begin{align}
    &\sum_j \int_{x_1}^{z_b}\! \frac{\df z_1}{z_1}\,\frac{f_j(x_1/z_1,\mu)}{f_q(x_1,\mu)}\, C_j P_{qj}(z_1)\,
    \ln \min \bigg\{\Big(\frac{p_1^-}{z_1 \mu}\Big)^2, \frac{\Tau\, p_1^-}{z_1(1-z_1) \mu^2}, \frac{p_T^2}{(1-z_1)^2\mu^2} \bigg\}
    \nn \\ & \quad
    = \sum_j \int_{x_1}^{1}\! \frac{\df z_1}{z_1}\,\frac{f_j(x_1/z_1,\mu)}{f_q(x_1,\mu)}\, C_j P_{qj}(z_1)\,
    \bigg[\ln \Big(\frac{p_T^2}{\mu^2} \Big) - 2 \ln(1-z_1)\bigg]
    \nn \\ & \qquad
    - \int_{z_b}^{1}\! \df z_1\, C_F P_{qq}(z_1)\,
    \bigg[\ln \Big(\frac{p_T^2}{\mu^2} \Big) - 2 \ln(1-z_1)\bigg] \Big[1+ \ORd{\frac{p_T^2}{\Tau Q}}\Big]
    \nn \\ & \quad
    = \sum_j \int_{x_1}^{1}\! \frac{\df z_1}{z_1}\,\frac{f_j(x_1/z_1,\mu)}{f_q(x_1,\mu)}\, C_j P_{qj}(z_1)\,
    \bigg[\ln \Big(\frac{p_T^2}{\mu^2} \Big) - 2 \ln(1-z_1)\bigg]
    \nn \\ & \qquad
   + C_F \bigg\{ 2 \ln^2 \Big(\frac{p_1^-}{\mu}\Big) 
       +4 \ln \Big(\frac{p_1^-}{\mu}\Big) \ln \Big(\frac{\Tau}{\mu}\Big)
       +2 \ln^2 \Big(\frac{\Tau}{\mu}\Big)
    \nn \\ & \qquad \quad
       - 2\ln \Big(\frac{p_T^2}{\mu^2}\Big) \bigg[\ln \Big(\frac{p_1^-}{\mu}\Big) + \ln \Big(\frac{\Tau}{\mu}\Big) + \frac34\bigg]
       \bigg\}  \Big[1+ \ORd{\frac{p_T^2}{\Tau Q}}\Big]
\,.\end{align}
We have verified that this agrees with the \SCETab factorization formula in \eq{DY_fact_SCETonetwo} expanded to NLO, providing an important check on our effective theory framework.

We now consider $z_b>z_a$, i.e.~$p_T < \Tau$, which is only allowed by the power counting in the \SCETb region of phase space. In contrast with \eq{min_regions}, we now only have two regions: $1 \geq z_1 \geq z_c$ and $z_c \geq z_1 \geq x_1$, where
\begin{equation}
  z_c = \frac{1}{1+ p_T/p_1^-}
\,.\end{equation}
This leads to the following correction to the \SCETab result,
\begin{align}
    &\theta(\Tau- p_T) \bigg\{\sum_j \int_{z_a}^{z_b}\! \frac{\df z_1}{z_1}\,\frac{f_j(x_1/z_1,\mu)}{f_q(x_1,\mu)}\,  C_j P_{qj}(z_1)\,
    \bigg[\ln \Big(\frac{\Tau\, p_1^-}{z_1(1-z_1) \mu^2} \Big) - \ln \Big(\frac{(p_1^-)^2}{z_1^2 \mu^2}\Big) \bigg]
    \nn \\
    &+ \sum_j \int_{z_b}^{z_c}\! \frac{\df z_1}{z_1}\,\frac{f_j(x_1/z_1,\mu)}{f_q(x_1,\mu)}\,  C_j P_{qj}(z_1)\,
    \bigg[ \ln \Big(\frac{p_T^2}{(1-z_1)^2\mu^2}\Big)- \ln \Big(\frac{(p_1^-)^2}{z_1^2 \mu^2}\Big) \bigg]\bigg\}
    \nn \\ & \quad
    = C_F\, \theta(\Tau - p_T)\Bigg\{ \int_{z_a}^{z_b}\! \df z_1\,  P_{qq}(z_1)  \bigg[\ln \Big(\frac{\Tau}{p_1^-}\Big) - \ln(1-z_1) \bigg] 
    \nn \\ & \qquad +
    \int_{z_b}^{z_c}\! \df z_1\,  P_{qq}(z_1)  \bigg[2\ln \Big(\frac{p_T}{p_1^-}\Big) -2 \ln(1-z_1) \bigg]  \bigg\} \Big[1+ \ORd{\frac{\Tau}{Q}}\Big]
    \nn \\ & \quad
    = \frac12\, C_F\, \theta(\Tau - p_T) \ln^2 \Big(\frac{p_T^2}{\Tau^2}\Big) \Big[1+ \ORd{\frac{\Tau}{Q}}\Big]
\,.\end{align}
The first line erases the earlier contributions from $z_a<z_1<z_b$ and the second line from $z_b<z_1<z_c$. This agrees with the FU soft function in \eq{scetb_matching}.

We conclude this section by briefly commenting on the size of the various power corrections we encountered. In \sec{IR_div}, we restricted to $p_T , \Tau \ll Q$, dropping some (but not all) terms of $\ord{p_T^2/Q^2, \Tau/Q}$. In our \SCETa  analysis in this section, we systematically expanded up to corrections of $\ord{\Tau/Q}$. For \SCETb the power corrections were $\ord{\Tau/Q \sim p_T^2/(\Tau Q)}$, and for \SCETab they were $\ord{p_T^2/(\Tau Q)}$ in size. Contrary to our expectation in \sec{scet}, we found no $\ord{\Tau^2/p_T^2}$ power corrections at NLO. However, it is quite possible that this changes at higher orders.

\section{Measuring Two Angularities on One Jet}
\label{sec:ang}

   \begin{table}
   \centering
   \begin{tabular}{l|cc}
     \hline \hline
     Mode: & Scaling $(- , + ,\perp)$  \\ \hline
     collinear & $Q(1,\lambda^{2r/\bt},\lambda^{r/\bt})$  \\
     collinear-soft & $ Q\Big(\lambda^{\frac{\al r - \bt }{\al-\bt}},\lambda^{\frac{(\al-2)r-(\bt-2)}{\al-\bt}},\lambda^{\frac{(\al-1)r-(\bt-1)}{\al-\bt}}\Big)$ \\ 
     soft & $ Q(\lambda,\lambda,\lambda)$  \\
     \hline \hline
   \end{tabular}
   \caption{Modes and power counting in \SCETab for the double angularity measurement on a single jet. The power counting parameter is $\lambda$, with
   $\la \sim e_\al \sim e_\bt^{1/r}$ and $1 > r > \bt/\al$.   }
   \label{tab:ang_modes}
   \end{table}

We will now apply our effective field theory framework to the measurement of two angularities on one jet. The angularity $e_\al$ of a jet is defined as~\cite{Berger:2003iw,Almeida:2008yp,Ellis:2010rwa}
\begin{align}
  e_\al = \sum_{i \in \text{jet}} \frac{E_i}{E_\text{jet}} \Big(\frac{\theta_i}{R}\Big)^\al
\,.\end{align}
Here, $E_i$ and $\theta_i$ denote the energy and angle (with respect to the jet axis) of particle $i$, and $E_\text{jet}$ and $R$ are the jet energy and radius. To avoid the issue of recoil \cite{Catani:1992jc,Dokshitzer:1998kz,Banfi:2004yd,Larkoski:2014uqa}, we use the winner-take-all axis \cite{Bertolini:2013iqa,Larkoski:2014uqa}. This ensures that the direction of the collinear radiation coincides with the jet axis. 

For the measurement of two angularities $e_\al$, $e_\bt$ (with $\al>\bt$), the phase space is given by $e_\al^{\bt/\al} \geq e_\bt \geq e_\al$ at NLL. The effective field theories on the boundaries were discussed in ref.~\cite{Larkoski:2014tva}, so we focus on the intermediate regime described by \SCETab. The modes of \SCETab are shown in table \ref{tab:ang_modes}.  Their power counting is fixed by the requirement that these modes are on-shell, that the collinear mode contributes to $e_\bt$, the soft mode contribute to $e_\al$ and the collinear-soft mode contribute to both. This leads to the following factorization formula 
\begin{align}\label{eq:ang_fact}
  \frac{\df^2\si_i}{\df e_\al\,\df e_\bt} &= \hat \si_i^\zero H_i(Q^2,\mu) \int \df e_\bt^\text{c}Q^\bt\, \df e_\al^\text{cs}Q\, \df e_\bt^\text{cs}Q^\bt\, \df e_\al^\text{s}Q\, J_i(e_\bt^\text{c} Q^\bt,\mu)\, \CS_i(e_\al^\text{cs}Q, e_\bt^\text{cs}Q^\bt)\, S_i(e_\al^\text{s} Q,\mu) 
  \nn \\ & \quad \times
  \de(e_\al  - e_\al^\text{cs} - e_\al^\text{c}) \de(e_\bt - e_\bt^\text{c} - e_\bt^\text{cs})
\,,\end{align}
for quark ($i=q$) and gluon ($i=g$) jets.
Here, $\hat{\si}_i^\zero$ is the tree-level cross section, and $H$ the hard function describing hard virtual corrections. The jet function $J$, soft function $S$ and collinear-soft function $\CS$ capture the effect of collinear, soft and collinear-soft radiation, respectively. The first two have been defined in ref.~\cite{Larkoski:2014tva} while the third is the analog of \eq{CSdef} but for the double angularity measurement. 
Since we only work up to NLL order, we are allowed to consider a single jet. At higher orders we need to take the rest of the event into account, and \eq{ang_fact} must accordingly be generalized to e.g.~$e^+e^-$ event shapes. We expect the power corrections to be $\ord{e_\al/e_\bt,e_\bt^{\al/\bt}/e_\al}$, which blow up at the edges of the phase space, where the boundary theories should be used instead.

Below we collect what is needed for NLL resummation. The RG equation and the anomalous dimension of the hard function are
\begin{align}
\mu \frac{\df}{\df\mu}\, H_i(Q^2, \mu) &= \gamma_H^i(Q^2, \mu)\, H_i(Q^2, \mu) 
\,, \nn \\
 \ga_H^i(Q^2,\mu) &= \Ga^i_\text{cusp}(\al_s) \ln \frac{Q^2}{\mu^2} + \ga_H^i(\al_s)
\,. 
\end{align}
For the jet function we have
\begin{align}
\mu \frac{\df}{\df\mu}\, J_i(e_\bt Q^{\bt} , \mu) &= \int_0^{e_\bt}\! \df {e_\bt'}\, Q^{\bt}\,  \ga_J^i(e_\bt Q^{\bt} - e_\bt' Q^{\bt},\mu)\, J_i(e_\bt' Q^{\bt}, \mu) 
\,, \nn \\
\ga_J^i(e_\bt Q^{\bt},\mu) &= -\frac{2}{\bt-1}\, \Ga^i_\text{cusp}(\al_s)\, \frac{1}{\mu^{\bt}} \cL_0\Big(\frac{e_\bt Q^{\bt}}{\mu^{\bt}}\Big) + \ga_J^i(\al_s)\,\de(e_\bt Q^{\bt})
\,,
\end{align}
and the soft function satisfies
\begin{align}
\mu \frac{\df}{\df\mu}\, S_i(e_\al Q, \mu) &= \int_0^{e_\al}\! \df {e_\al'} Q\,  \ga_S^i(e_\al Q - e_\al' Q,\mu)\, S_i(e_\al' Q, \mu) 
\,, \nn \\
\ga_S^i(e_\al Q,\mu) &= \frac{2}{\al-1}\, \Ga^i_\text{cusp}(\al_s)\, \frac{1}{\mu} \cL_0\Big(\frac{e_\al Q}{\mu}\Big) + \ga_S^i(\al_s)\,\de(e_\al Q)
\,.\end{align}
   The anomalous dimension of the collinear-soft function is constrained by consistency of the cross section in \eq{ang_fact}. 
  These anomalous dimensions involve $\Ga^i_\text{cusp}(\al_s)$, given in \app{RGE}, and the non-cusp parts
  \begin{align}
    \ga^i_X(\al_s) = \sum_n \ga^i_{X,n} \Big(\frac{\al_s}{4\pi}\Big)^{n+1}
  \,,\end{align}
  with $X=H,J,S$. At NLL we only need the leading coefficients,
\begin{align} \label{eq:ga_coeff}
 \ga_{H,0}^q  = -6 C_F
  \,, \qquad
  \ga_{H,0}^g  = -2 \beta_0 
  \,, \qquad
 \ga_{J,0}^i  = - \ga_{H,0}^i
  \,,  \qquad
  \ga_{S,0}^i = 0
\,,\end{align}
where $\beta_0 = \frac{11}{3}\,C_A -\frac{4}{3}\,T_F\,n_f$.

We now evaluate the double cumulative distribution at NLL order by inserting the tree-level expressions
\begin{align}
  H_i(Q^2,\mu) &= 1
  \,, &
  J_i(e_\bt Q^\bt,\mu) &= \de(e_\bt Q^\bt)
  \,, \nn \\
  \CS_i(e_\al Q, e_\bt Q^\bt) &= \de(e_\al Q)\, \de(e_\bt Q^\bt)
  \,, &
  S_i(e_\al Q,\mu) &= \de(e_\al Q)
\,,\end{align}
 in \eq{ang_fact} and evolving them to the collinear-soft scale $\mu_\CS$. This results in
\begin{align} \label{eq:nll}
\Sigma_i(e_\al,e_\bt) &= \int_0^{e_\al} \! \df e_\al'\, \int_0^{e_\bt} \! \df e_\bt'\, \frac{\partial^2 \si}{\partial e_\al' \partial e_\bt'} 
\nn \\ &
= \hat \si_i^\zero\, \frac{e^{K_H^i + K_J^i + K_S^i - \gamma_E\, \eta_J^i - \gamma_E\, \eta_S^i}}{\Gamma(1+\eta_J^i)\Gamma(1+\eta_S^i)}\, 
 \Bigl(\frac{Q}{\mu_H}\Bigr)^{2\eta_H^i}
\Bigl( \frac{e_\bt^{1/\beta} Q}{\mu_J} \Bigr)^{\beta\, \eta_J^i}
\Bigl( \frac{e_\al Q}{\mu_S} \Bigr)^{\eta_S^i}
\,.\end{align}
The evolution kernels that enter here are
\begin{align}
K_H^i(\mu_H,\mu_\CS) &= -2 K_\Gamma^i(\mu_H,\mu_\CS) + K_{\gamma_H^i}(\mu_H,\mu_\CS)
\,, \quad &
\eta_H^i(\mu_J,\mu_\CS) &= \eta_{\Gamma}^i(\mu_J,\mu_\CS)
\,,\nn \\
K_J^i(\mu_J,\mu_\CS) &= -\frac{2 \beta}{1-\beta}\, K_\Gamma^i(\mu_J,\mu_\CS) + K_{\gamma_J^i}(\mu_J,\mu_\CS)
\,, \quad &
\eta_J^i(\mu_J,\mu_\CS) &= \frac{2}{1-\beta}\, \eta_{\Gamma}^i(\mu_J,\mu_\CS)
\,,\nn\\
K_S^i(\mu_S,\mu_\CS) &= \frac{2}{1-\al}\, K_\Gamma^i(\mu_S,\mu_\CS)
\,, \quad &
\eta_S^i(\mu_S,\mu_\CS) &= -\frac{2}{1-\al}\, \eta_{\Gamma}^i(\mu_S,\mu_\CS)
\,,\end{align}
in terms of $K_\Ga^i$, $\eta_\Ga^i$ and $K_{\ga_X^i}$ defined in \eq{Keta_def}.
As starting point for the RG evolution we use the canonical (natural) scales
  \begin{align} \label{eq:ang_scales}
    \mu_H &= Q
    \,, \nn \\    
    \mu_J &= e_\bt^{1/\bt} Q = \mu_{J \to J}  
    \,, \nn \\
    \mu_\CS &= \big(e_\al^{1-\bt} e_\bt^{\al-1}\big)^{1/(\al-\bt)} Q = \mu_{J \to S}
    \,, \nn \\
    \mu_S &= e_\al Q = \mu_{S \to S}
  \,.\end{align}
  which we identified with the interpolating scales $\mu_{J \to J}$, $\mu_{J \to S}$ and $\mu_{S \to S}$ of ref.~\cite{Larkoski:2014tva} (see also app. C of ref.~\cite{Larkoski:2014pca}) to simplify the comparison.

This mostly agrees with the conjecture made in ref.~\cite{Larkoski:2014tva} 
\begin{align} \label{eq:AID}
\Sigma^\text{ref.\cite{Larkoski:2014tva}}_i(e_\al,e_\bt) 
& = \frac{e^{- R(e_\al,e_\bt) - \gamma_E\, \tilde R(e_\al,e_\bt)}}{\Gamma(1+ \tilde R(e_\al,e_\bt))}\, 
\end{align}
where
\begin{align}
R(e_\al,e_\bt) &\stackrel{\text{NLL}}{=} -K_H^i(\mu_H,\mu_\CS) - K_J^i(\mu_J,\mu_\CS) - K_S^i(\mu_S,\mu_\CS) 
 \,, \nn \\
\tilde R(e_\al,e_\bt) &\stackrel{\text{NLL}}{=} \eta_J^i(\mu_J,\mu_\CS) + \eta_S^i(\mu_S,\mu_\CS)
 \,.\end{align}
The only difference\footnote{Ignoring differences beyond NLL order and power suppressed contributions.} with our result in \eq{nll} is in the denominator, where we have  $\Gamma(1+\eta_J^i)\Gamma(1+\eta_S^i)$ instead of $\Gamma(1+\eta_J^i+\eta_S^i)$. These factors agree with each other on the boundary, since either $\eta_J^i$ or $\eta_S^i$ vanishes there, but lead to $\ord{\al_s^2 \ln^2}$ differences in the bulk. (An analogous conjecture to \eq{AID} in Laplace space does agree with our result.\footnote{We thank D. Neill for pointing this out.})

According to ref.~\cite{Larkoski:2014tva}, the leading difference between their interpolation and the true NLL cross section
is expected to be $\al_s^4 \ln^4$. However, this is based on boundary conditions for the differential cross section, which do not affect the logarithmic accuracy of their calculation in the bulk. Specifically,  their differential cross section satisfies the condition at the boundary $e_\al = e_\bt^{\al/\bt}$ through the addition of terms that are power suppressed. Since these terms are power suppressed in the bulk, they cannot improve the logarithmic accuracy there. 

In ref.~\cite{numerics}, we will discuss how a more sophisticated scale choice than \eq{ang_scales} provides a natural way to satisfy the derivative boundary condition. In addition to requiring $\mu_\CS$ to merge with $\mu_J$ or $\mu_S$ on the respective boundaries, one can also require a continuous derivative,
\begin{align}\label{eq:deriv_constr}
  \frac{\partial}{\partial e_\al}\, \mu_J(e_\al,e_\bt) \Big|_{e_\bt = e_\al^{\bt/\al}} &= \frac{\df}{\df e_\al}\, \mu_J(e_\al,  e_\al^{\bt/\al})
 \,, &
  \frac{\partial}{\partial e_\bt}\, \mu_J(e_\al,e_\bt) \Big|_{e_\bt = e_\al^{\bt/\al}} &= 0
  \,,\nn \\
  \frac{\partial}{\partial e_\al}\, \mu_\CS(e_\al,e_\bt) \Big|_{e_\bt = e_\al^{\bt/\al}} &= 0
 \,, &
  \frac{\partial}{\partial e_\bt}\, \mu_\CS(e_\al,e_\bt) \Big|_{e_\bt = e_\al^{\bt/\al}} &= 0
  \,,\nn \\
  \frac{\partial}{\partial e_\al}\, \mu_S(e_\al,e_\bt) \Big|_{e_\bt = e_\al^{\bt/\al}} &= \frac{\df}{\df e_\al}\, \mu_S(e_\al,  e_\al^{\bt/\al})
 \,, &
  \frac{\partial}{\partial e_\bt}\, \mu_S(e_\al,e_\bt) \Big|_{e_\bt = e_\al^{\bt/\al}} &= 0
\,,\end{align}
and similarly for the boundary at $e_\al=e_\bt$. These equations closely resemble those imposed on $R$ and $\tilde R$ in ref.~\cite{Larkoski:2014tva} and follow from the same steps.
Note that there is a redundancy in the constraints in \eq{deriv_constr}, as e.g.~the second equation on the first line implies the first.
The scale choice in transitioning to a region where resummation is turned off has been studied for single variables in e.g.~refs.~\cite{Ligeti:2008ac,Abbate:2010xh}, and also in ref.~\cite{Bauer:2011uc}.   
  
\section{Conclusions}
\label{sec:conclusions}

In this paper we studied the resummation of double differential measurements. We focussed on two examples: Drell-Yan production with a (beam-thrust) jet veto where the $p_T$ of the lepton pair is measured, and the measurement of two angularities on one jet. Concerning the latter, in ref.~\cite{Larkoski:2014tva} resummation on the two phase space boundaries was achieved, and an interpolation was built to smoothly connect them. We go beyond this by identifying the factorization formula needed to achieve resummation in the intermediate regime. This involves additional collinear-soft modes, and the corresponding collinear-soft function was calculated at one loop. The relations between FU PDFs, collinear-soft functions and (FU) soft functions were investigated. The consistency of our factorization theorem was verified by checking that the anomalous dimensions cancel between the various ingredients, and by comparing to an analytic NLO calculation of the cross section. We also showed how to combine the factorization theorems on the boundaries and interior, to achieve NNLL precision throughout. At variance with ref.~\cite{Larkoski:2014tva} we found a universal factorization formula that describes the cross section in all three phase space regions up to power corrections. Numerical results, including the matching to fixed order, will be presented in ref.~\cite{numerics}.

If the hierarchy of scales for the individual variables is not that large, such that the resummation of them is only marginally important, there may be not enough room for a distinct \SCETab region of phase space. (This can be seen in \fig{regions}, where you have to go deeper into the resummation region for \SCETab.) Even in this case, one benefits from knowing the correct description of the intermediate regime in building the interpolation between boundaries, as illustrated by the $\ord{\al_s^2 \ln^2}$ difference between our NLL results and the interpolation conjectured in ref.~\cite{Larkoski:2014tva}.

Looking forward, we hope the results presented here will stimulate the development of more realistic analytic resummations and more robust Monte Carlo descriptions of LHC events. The framework presented here has natural generalizations to resummation in more than two variables. Finally, finding a proper description of the ``terra incognita" in \fig{regions} is important for resolving a long-standing issue over double counting between higher-order corrections and double parton scatterings.

\begin{acknowledgments}
  We thank Thomas Becher and Mathias Ritzmann for discussions. 
  We thank Andrew Larkoski, Ian Moult, Duff Neil, Frank Tackmann and Jonathan Walsh for feedback on this manuscript. 
  W.W. thanks the Galileo Galilei Institute for Theoretical Physics for hospitality and the INFN for partial support during the completion of this work. 
  M.P. acknowledges support by the Swiss National Science Foundation.
  W.W. is supported by a Marie Curie International Incoming Fellowship within the 7th European Community Framework Program (PIIF-GA-2012-328913).
 This work is part of the D-ITP consortium, a program of the Netherlands Organization for Scientific Research (NWO) that is funded by the Dutch Ministry of Education, Culture and Science (OCW).
\end{acknowledgments}

\appendix

\section{Plus Distributions}
\label{app:plusdist}

The standard plus distribution for some function $g(x)$ can be defined as
\begin{equation} \label{eq:plus_def}
\bigl[\theta(x) g(x)\bigr]_+
= \lim_{\beta \to 0} \frac{\df}{\df x} \bigl[\theta(x-\beta)\, G(x) \bigr]
\qquad\text{with}\qquad
G(x) = \int_1^x\!\df x'\, g(x')
\,,\end{equation}
satisfying the boundary condition $\int_0^1 \df x\, [\theta(x) g(x)]_+ = 0$. Two special cases we need are
\begin{align} \label{eq:plusdef}
\cL_n(x)
&\equiv \biggl[ \frac{\theta(x) \ln^n x}{x}\biggr]_+
 = \lim_{\beta \to 0} \biggl[
  \frac{\theta(x- \beta)\ln^n x}{x} +
  \delta(x- \beta) \, \frac{\ln^{n+1}\!\beta}{n+1} \biggr]
\,,\nn\\
\cL^\eta(x)
&\equiv \biggl[ \frac{\theta(x)}{x^{1-\eta}}\biggr]_+
 = \lim_{\beta \to 0} \biggl[
  \frac{\theta(x - \beta)}{x^{1-\eta}} +
  \delta(x- \beta) \, \frac{x^\eta - 1}{\eta} \biggr]
\,.\end{align}
In our calculations, we use the following expansion in plus distributions
\begin{equation} \label{eq:plusexp}
 \frac{\theta(x)}{x^{1+\eps}} = - \frac{1}{\eps}\,\delta(x) + \cL_0(x)
  - \eps \cL_1(x) + \ord{\eps^2}
\,.\end{equation}
Rescaling and convolution identities for $\cL_n(x)$ and $\cL^\eta(x)$ can be found in App.~B of ref.~\cite{Ligeti:2008ac}.

\section{Renormalization Group Evolution}
\label{app:RGE}

The functions $K_\Gamma^i(\mu_0, \mu)$, $\eta_\Gamma^i(\mu_0, \mu)$ and $K_{\gamma_X^i}(\mu_0, \mu)$ that enter in the RGE solutions are defined by
\begin{align} \label{eq:Keta_def}
K_\Gamma^i(\mu_0, \mu)
& = \int_{\alpha_s(\mu_0)}^{\alpha_s(\mu)}\!\frac{\df\alpha_s}{\beta(\alpha_s)}\,
\Gamma_\cusp^i(\alpha_s) \int_{\alpha_s(\mu_0)}^{\alpha_s} \frac{\df \alpha_s'}{\beta(\alpha_s')}
\,,\qquad
\eta_\Gamma^i(\mu_0, \mu)
= \int_{\alpha_s(\mu_0)}^{\alpha_s(\mu)}\!\frac{\df\alpha_s}{\beta(\alpha_s)}\, \Gamma_\cusp^i(\alpha_s)
\,,\nn \\
K_{\gamma_X^i}(\mu_0, \mu)
& = \int_{\alpha_s(\mu_0)}^{\alpha_s(\mu)}\!\frac{\df\alpha_s}{\beta(\alpha_s)}\, \gamma_X^i(\alpha_s)
\,.\end{align}
Expanding the beta function and anomalous dimensions in powers of $\alpha_s$,
\begin{align}
\beta(\alpha_s) &=
- 2 \alpha_s \sum_{n=0}^\infty \beta_n\Bigl(\frac{\alpha_s}{4\pi}\Bigr)^{n+1}
\,, \quad
\Gamma_\cusp^i(\alpha_s) = \sum_{n=0}^\infty \Gamma_n^i \Bigl(\frac{\alpha_s}{4\pi}\Bigr)^{n+1}
\,, \quad
\gamma_X^i(\alpha_s) = \sum_{n=0}^\infty \gamma_{X,n}^i \Bigl(\frac{\alpha_s}{4\pi}\Bigr)^{n+1}
\,,\end{align}
their explicit expressions at NNLL order are
\begin{align} \label{eq:Keta}
K_\Gamma^i(\mu_0, \mu) &= -\frac{\Gamma^i_0}{4\beta_0^2}\,
\biggl\{ \frac{4\pi}{\alpha_s(\mu_0)}\, \Bigl(1 - \frac{1}{r} - \ln r\Bigr)
   + \biggl(\frac{\Gamma^i_1 }{\Gamma^i_0 } - \frac{\beta_1}{\beta_0}\biggr) (1-r+\ln r)
   + \frac{\beta_1}{2\beta_0} \ln^2 r
\nn\\ & \hspace{10ex}
+ \frac{\alpha_s(\mu_0)}{4\pi}\, \biggl[
  \biggl(\frac{\beta_1^2}{\beta_0^2} - \frac{\beta_2}{\beta_0} \biggr) \Bigl(\frac{1 - r^2}{2} + \ln r\Bigr)
  + \biggl(\frac{\beta_1\Gamma^i_1 }{\beta_0 \Gamma^i_0 } - \frac{\beta_1^2}{\beta_0^2} \biggr) (1- r+ r\ln r)
\nn\\ & \hspace{10ex}
  - \biggl(\frac{\Gamma^i_2 }{\Gamma^i_0} - \frac{\beta_1\Gamma^i_1}{\beta_0\Gamma^i_0} \biggr) \frac{(1- r)^2}{2}
     \biggr] \biggr\}
\,, \nn\\
\eta_\Gamma^i(\mu_0, \mu) &=
 - \frac{\Gamma^i_0}{2\beta_0}\, \biggl[ \ln r
 + \frac{\alpha_s(\mu_0)}{4\pi}\, \biggl(\frac{\Gamma^i_1 }{\Gamma^i_0 }
 - \frac{\beta_1}{\beta_0}\biggr)(r\!-\!1)
 + \frac{\alpha_s^2(\mu_0)}{16\pi^2} \biggl(
    \frac{\Gamma^i_2 }{\Gamma^i_0 } - \frac{\beta_1\Gamma^i_1 }{\beta_0 \Gamma^i_0 }
      + \frac{\beta_1^2}{\beta_0^2} -\frac{\beta_2}{\beta_0} \biggr) \frac{r^2\!-\!1}{2}
    \biggr]
\,, \nn\\
K_{\gamma_X^i}(\mu_0, \mu) &=
 - \frac{\gamma_{X,0}^i}{2\beta_0}\, \biggl[ \ln r
 + \frac{\alpha_s(\mu_0)}{4\pi}\, \biggl(\frac{\gamma_{X,1}^i }{\gamma_{X,0}^i }
 - \frac{\beta_1}{\beta_0}\biggr)(r-1) \biggr]
\,.\end{align}
Here, $r = \alpha_s(\mu)/\alpha_s(\mu_0)$ and the running coupling is given by the three-loop expression
\begin{equation} \label{eq:alphas}
\frac{1}{\alpha_s(\mu)} = \frac{X}{\alpha_s(\mu_0)}
  +\frac{\beta_1}{4\pi\beta_0}  \ln X
  + \frac{\alpha_s(\mu_0)}{16\pi^2} \biggr[
  \frac{\beta_2}{\beta_0} \Bigl(1-\frac{1}{X}\Bigr)
  + \frac{\beta_1^2}{\beta_0^2} \Bigl( \frac{\ln X}{X} +\frac{1}{X} -1\Bigr) \biggl]
\,,\end{equation}
where $X = 1+\alpha_s(\mu_0)\beta_0 \ln(\mu/\mu_0)/(2\pi)$.

The coefficients of the beta function~\cite{Tarasov:1980au, Larin:1993tp}, cusp anomalous dimension~\cite{Korchemsky:1987wg, Moch:2004pa}, non-cusp anomalous dimensions of the hard function and jet function~\cite{Moch:2004pa,Moch:2005id,Moch:2005tm,Vogt:2004mw,Idilbi:2005ni,Idilbi:2006dg,Becher:2006mr,Becher:2009th,Berger:2010xi} and non-cusp anomalous dimension of the rapidity resummation~\cite{Davies:1984sp,deFlorian:2001zd,Becher:2010tm,Gehrmann:2014yya} are given below in the $\overline{\mathrm{MS}}$ scheme. At this order $\Ga_i^g = (C_A/C_F) \Ga_i^q$, which are therefore not separately shown.
\begin{align} 
\beta_0 &= \frac{11}{3}\,C_A -\frac{4}{3}\,T_F\,n_f
\,,\nn\\
\beta_1 &= \frac{34}{3}\,C_A^2  - \Bigl(\frac{20}{3}\,C_A\, + 4 C_F\Bigr)\, T_F\,n_f
\,, \nn\\
\beta_2 &=
\frac{2857}{54}\,C_A^3 + \Bigl(C_F^2 - \frac{205}{18}\,C_F C_A
 - \frac{1415}{54}\,C_A^2 \Bigr)\, 2T_F\,n_f
 + \Bigl(\frac{11}{9}\, C_F + \frac{79}{54}\, C_A \Bigr)\, 4T_F^2\,n_f^2
\,, \\[2ex]
\Gamma_0^q &= 4 C_F
\,,\nn\\
\Gamma_1^q &= 4 C_F \Bigl[\Bigl( \frac{67}{9} -\frac{\pi^2}{3} \Bigr)\,C_A  -
   \frac{20}{9}\,T_F\, n_f \Bigr]
\,,\nn\\
\Gamma_2^q &= 4 C_F \Bigl[
\Bigl(\frac{245}{6} -\frac{134 \pi^2}{27} + \frac{11 \pi ^4}{45}
  + \frac{22 \zeta_3}{3}\Bigr)C_A^2
  + \Bigl(- \frac{418}{27} + \frac{40 \pi^2}{27}  - \frac{56 \zeta_3}{3} \Bigr)C_A\, T_F\,n_f
\nn\\* & \hspace{8ex}
  + \Bigl(- \frac{55}{3} + 16 \zeta_3 \Bigr) C_F\, T_F\,n_f
  - \frac{16}{27}\,T_F^2\, n_f^2 \Bigr]
\,, \\[2ex]
\gamma^q_{H\,0} &= -6 C_F
\,,\nn\\
\gamma^q_{H\,1}
&= - C_F \Bigl[
  \Bigl(\frac{82}{9} - 52 \zeta_3\Bigr) C_A
+ (3 - 4 \pi^2 + 48 \zeta_3) C_F
+ \Bigl(\frac{65}{9} + \pi^2 \Bigr) \beta_0 \Bigr]
\,,\nn\\
\gamma^q_{H\,2}
&= -2C_F \Bigl[
  \Bigl(\frac{66167}{324} - \frac{686 \pi^2}{81} - \frac{302 \pi^4}{135} - \frac{782 \zeta_3}{9} + \frac{44\pi^2 \zeta_3}{9} + 136 \zeta_5\Bigr) C_A^2
\nn\\ & \qquad\hspace{6ex}
+ \Bigl(\frac{151}{4} - \frac{205 \pi^2}{9} - \frac{247 \pi^4}{135} + \frac{844 \zeta_3}{3} + \frac{8 \pi^2 \zeta_3}{3} + 120 \zeta_5\Bigr) C_F C_A
\nn\\ & \qquad\hspace{6ex}
+ \Bigl(\frac{29}{2} + 3 \pi^2 + \frac{8\pi^4}{5} + 68 \zeta_3 - \frac{16\pi^2 \zeta_3}{3} - 240 \zeta_5\Bigr) C_F^2
\nn\\ & \qquad\hspace{6ex}
+ \Bigl(-\frac{10781}{108} + \frac{446 \pi^2}{81} + \frac{449 \pi^4}{270} - \frac{1166 \zeta_3}{9} \Bigr) C_A \beta_0
\nn\\ & \qquad\hspace{6ex}
+ \Bigl(\frac{2953}{108} - \frac{13 \pi^2}{18} - \frac{7 \pi^4 }{27} + \frac{128 \zeta_3}{9}\Bigr)\beta_1
+ \Bigl(-\frac{2417}{324} + \frac{5 \pi^2}{6} + \frac{2 \zeta_3}{3}\Bigr)\beta_0^2
\Bigr]
\,,\\[2ex]
\gamma_{H\,0}^g &= -2 \beta_0
\,,\nn\\
\gamma_{H\,1}^g
&= \Bigl(-\frac{118}{9} + 4\zeta_3\Bigr)C_A^2 +
\Bigl(-\frac{38}{9}+\frac{\pi^2}{3} \Bigr) C_A\, \beta_0 - 2 \beta_1
\,,\nn\\
\gamma_{H\,2}^g
&= \Bigl(-\frac{60875}{162} + \frac{634\pi^2}{81} +\frac{8\pi^4}{5}+\frac{1972\zeta_3}{9}
- \frac{40\pi^2 \zeta_3}{9} - 32\zeta_5 \Bigr)C_A^3
\nn \\ & \quad +
\Bigl(\frac{7649}{54}+\frac{134\pi^2}{81} - \frac{61\pi^4}{45}
 - \frac{500\zeta_3}{9}\Bigr) C_A^2\, \beta_0 +
\Bigl(\frac{466}{81}+\frac{5\pi^2}{9}-\frac{28 \zeta_3}{3}\Bigr) C_A\, \beta_0^2
\nn \\ & \quad +
\Bigl(-\frac{1819}{54} + \frac{\pi^2}{3} + \frac{4\pi^4}{45} + \frac{152\zeta_3}{9}\Bigr) C_A\, \beta_1
-2 \beta_2
\,,\\[2ex]
\gamma_{J\,0}^q &= 6 C_F
\,,\nn\\
\gamma_{J\,1}^q
&= C_F \Bigl[
  \Bigl(\frac{146}{9} - 80 \zeta_3\Bigr) C_A
+ (3 - 4 \pi^2 + 48 \zeta_3) C_F
+ \Bigl(\frac{121}{9} + \frac{2\pi^2}{3} \Bigr) \beta_0 \Bigr]
\,,\nn\\
\gamma_{J\,2}^q
&= 2 C_F \Bigl[
  \Bigl(\frac{52019}{162} - \frac{841\pi^2}{81} - \frac{82\pi^4}{27} -\frac{2056\zeta_3}{9} + \frac{88\pi^2 \zeta_3}{9} + 232 \zeta_5\Bigr)C_A^2
\nn\\ & \quad\hspace{6ex}
+ \Bigl(\frac{151}{4} - \frac{205\pi^2}{9} - \frac{247\pi^4}{135} + \frac{844\zeta_3}{3} + \frac{8\pi^2 \zeta_3}{3} + 120 \zeta_5\Bigr) C_A C_F
\nn\\ & \quad\hspace{6ex}
+ \Bigl(\frac{29}{2} + 3 \pi^2 + \frac{8\pi^4}{5} + 68 \zeta_3 - \frac{16\pi^2 \zeta_3}{3} - 240 \zeta_5\Bigr) C_F^2
\nn\\ & \quad\hspace{6ex}
+ \Bigl(-\frac{7739}{54} + \frac{325}{81} \pi^2 + \frac{617 \pi^4}{270} - \frac{1276\zeta_3}{9} \Bigr) C_A\beta_0
\nn\\ & \quad\hspace{6ex}
+ \Bigl(-\frac{3457}{324} + \frac{5\pi^2}{9} + \frac{16 \zeta_3}{3} \Bigr) \beta_0^2
+ \Bigl(\frac{1166}{27} - \frac{8 \pi^2}{9} - \frac{41 \pi^4}{135} + \frac{52 \zeta_3}{9}\Bigr) \beta_1
\Bigr]
\,,\\[2ex]
\gamma_{J\,0}^g &= 2 \beta_0
\,,\nn\\
\gamma_{J\,1}^g
&= \Bigl(\frac{182}{9} - 32\zeta_3\Bigr)C_A^2 +
\Bigl(\frac{94}{9}-\frac{2\pi^2}{3}\Bigr) C_A\, \beta_0 + 2\beta_1
\,,\nn\\
\gamma_{J\,2}^g
&= \Bigl(\frac{49373}{81} - \frac{944 \pi^2}{81} - \frac{16\pi^4}{5} - \frac{4520 \zeta_3}{9}
+ \frac{128\pi^2 \zeta_3}{9} + 224 \zeta_5 \Bigr)C_A^3
\nn \\ & \quad +
\Bigl(-\frac{6173}{27} - \frac{376 \pi^2}{81} + \frac{13\pi^4}{5}
+ \frac{280\zeta_3}{9}\Bigr) C_A^2\, \beta_0 +
\Bigl(-\frac{986}{81}-\frac{10\pi^2}{9}+\frac{56 \zeta_3}{3}\Bigr) C_A\, \beta_0^2
\nn \\ & \quad +
\Bigl(\frac{1765}{27} - \frac{2\pi^2}{3} - \frac{8\pi^4}{45} - \frac{304\zeta_3}{9}\Bigr) C_A\, \beta_1
+ 2 \beta_2
\,,\\[2ex]
\gamma_{\nu,0}^q &= 0
\,,\nn \\
\gamma_{\nu,1}^q &= C_F \Big[\Big(\frac{64}{9} - 28 \zeta_3\Big) C_A + 32\zeta_3\, C_F + \frac{56}{9}\, \beta_0\Big]
\,,\\[2ex]
\gamma_{\nu,0}^g &= 0
\,,\nn \\
\gamma_{\nu,1}^g &= C_A \Big[\Big(\frac{64}{9} +4 \zeta_3\Big) C_A + \frac{56}{9}\, \beta_0\Big]
\,. \end{align}

\bibliographystyle{jhep}
\bibliography{scet1_5}


\end{document}